\documentclass[12pt]{emulateapj} 
\usepackage{amsmath}
\usepackage{amssymb}
\usepackage{amstext}
\usepackage{graphicx}
\usepackage{epstopdf}
\usepackage{epsfig}
\usepackage{color}
\definecolor{myColor}{rgb}{0.9,0.9,0.9}  
\begin{document}
\renewcommand\bottomfraction{.9}
\title{Analytic models for albedos, phase curves and polarization of reflected light from exoplanets}
\author{Nikku Madhusudhan\altaffilmark{1} and Adam Burrows\altaffilmark{1}}
\altaffiltext{1}{Department of Astrophysical Sciences, Princeton University, 
Princeton, NJ 08544 {\tt nmadhu@astro.princeton.edu, burrows@astro.princeton.edu}} 

\begin{abstract}
New observational facilities are becoming increasingly capable of observing reflected light 
from transiting and directly imaged extrasolar planets. In this study, we provide an analytic 
framework to interpret observed phase curves, geometric albedos, and polarization of giant planet 
atmospheres. We compute the observables for non-conservative Rayleigh scattering in homogeneous 
semi-infinite atmospheres using both scalar and vector formalisms. In addition, we compute phase curves and 
albedos for Lambertian, isotropic, and anisotropic scattering phase functions. We provide analytic expressions for 
geometric albedos and spherical albedos as a function of the scattering albedo for Rayleigh scattering in 
semi-infinite atmospheres. Given an observed geometric albedo our prescriptions can be used to estimate 
the underlying scattering albedo of the atmosphere, which in turn is indicative of the scattering and absorptive 
properties of the atmosphere. We also study the dependence of polarization in Rayleigh scattering atmospheres 
on the orbital parameters of the planet-star system, particularly on the orbital inclination. We show how the orbital 
inclination of non-transiting exoplanets can be constrained from their observed polarization parameters. 
We consolidate the formalism, solution techniques, and results from analytic models available in the literature, 
often scattered in various sources, and present a systematic procedure to compute albedos, phase curves, and 
polarization of reflected light. 
\end{abstract}


\section{Introduction}

Observations of reflected light provide constraints on the scattering and absorption 
processes in planetary atmospheres. In the solar system, geometric albedo spectra and 
phase curves have been used to infer chemical compositions and cloud properties in several 
planetary and satellite atmospheres (Hansen \& Hovenier 1974; Karkoschka 1994; Satoh et al. 2000; 
Sromovsky et al. 2001; Irwin et al. 2002; Karkoschka \& Tomasko 2011). Observational techniques are 
becoming increasingly capable of detecting reflected light from exoplanetary atmospheres, though currently 
available data are often limited to broadband visible photometry (Snellen et al. 2009, Borucki et al. 2010, 
Demory et al. 2011). Our goal in the present work is to provide an analytic framework to aid in the interpretation 
of measurements of geometric albedos and phase curves of irradiated exoplanets. 

Several theoretical studies in the past decade have advocated the use of reflected light observations for 
atmospheric characterization of extrasolar giant planets. Numerical models have been reported for directly 
imaged planets on wide orbits, as well as for closer-in giant planets with very high incident stellar irradiation 
(Seager \& Sasselov 1998; Marley et al. 1999; Seager et al. 2000;  Goukenleuque et al. 1999; Sudarsky et al. 2000,2005; Stam et al. 2004; Sengupta \& Maiti 2006; Burrows et al. 2008; Buenzli \& Schmid 2009; Cahoy et al. 2011; de Kok et al. 2011; Kostogryz et al. 2011; Marley \& Sengupta 2011). Atmospheres of extrasolar planets known to date, predominantly of extrasolar giant planets, experience a much wider range of equilibrium temperatures ($T_{\rm eq}  \sim$ 500 - 3000 K) than that of solar-system giant planets ($T_{\rm eq} \lesssim$ 125 K). Sudarsky et al. (2000) suggested a classification scheme of albedo spectra for giant planets, based on which highly irradiated planets ($T_{\rm eq} \sim$ 900 - 1500 K) are expected to have lower optical albedos due to strong Na and K absorption. At even higher temperatures, the presence of silicate condensates might increase the albedos. Albedo-based classification schemes for exoplanetary atmospheres were also discussed by  Burrows et al. (2008), Cahoy et al. (2011), and Cowan \& Agol (2010) as a function of different system parameters. 

Recently, observational constraints on the geometric albedos, visible phase curves, and potential polarization 
parameters, for several exoplanetary atmospheres have been reported. 
Some of the earliest constraints on geometric albedos ($A_g$) of extrasolar giant planets were 
reported for the non-transiting planets $\tau$ Boo, with upper limits of 0.3 (Charbonneau et al. 1999),  
0.22 (Collier-Cameron et al. 2000) and 0.39 (Leigh et al. 2003a), $\upsilon$ And with an upper limit of 0.42  
(Collier-Cameron et al. 2002), and HD75289 with an upper limit of 0.12 (Leigh et al. 2003b). However, in 
recent years, with the discoveries of transiting exoplanets and with the advent of high-precision transit photometry, 
geometric albedo constraints have been obtained for a substantial number of exoplanets. Rowe et al. (2008) observed the transiting hot Jupiter HD 209458b with the {\it MOST} 
satellite and reported a 3-sigma upper-limit of $A_g < 0.17$. Other hot Jupiters with $A_g$ estimates 
include TrES-3 (Winn et al. 2008), CoRoT-1b (Snellen et al. 2009), CoRoT-2b (Snellen et al. 2010), 
HAT-P-7b (Christiansen et al. 2010), Kepler-5b and 6b (Kipping \& Bakos 2010; Desert et al. 2011), 
HD~189733b (Berdyugina et al. 2011), and Kipping \& Spiegel (2011), all of which have $A_g$ upper-limits 
less than 0.3. On the other hand, Kipping \& Bakos (2010) and Demory et al. (2011) reported independent detections of a high geometric albedo for Kepler-7b of $A_g = 0.38 \pm 0.12$ and  $A_g = 0.32 \pm 0.03$. 
Recently, several groups have also reported attempts to measure polarization parameters of exoplanetary 
atmospheres in the visible (Berdyugina et al. 2008, 2011a,b; Wiktorowicz 2009).

In this work, we derive an analytic framework with which to interpret observations of albedos, phase curves, 
and polarization of exoplanetary atmospheres. We generally assume cloud-free\footnote{We do consider the asymmetric scattering phase function (see Section~\ref{sec:asymm}) which may be used as an approximation for particulate scattering if the scattering asymmetry factor is known. However, such a treatment still does not address the full problem of Mie scattering by particulates.} homogeneous and semi-infinite atmospheres, and analytically solve the multiple scattering problem for several different scattering phase functions. Detailed numerical models solve the general radiative transfer problem in a plane-parallel inhomogeneous atmosphere (Marley et al. 1999; Seager et al. 2000; Sudarsky et al. 2000; Burrows et al. 2008), with the assumption of radiative equilibrium for given chemistry and sources of absorption and scattering. Such models, however, involve significant computation time and convergence monitoring, and may be computationally prohibitive for formal fits to data, where several consecutive model evaluations are typically required. On the other hand, computationally efficient analytic models exist for several forms of scattering, if the atmosphere can be assumed to be homogeneous 
in the scattering albedo\footnote{The scattering albedo is given by $\omega = \sigma_{scat}/(\sigma_{scat} + \sigma_{abs})$, where $\sigma_{scat}$ is the single-scattering cross section and $\sigma_{abs}$ is the absorption cross section.} ($\omega$). While the latter condition may not be exactly satisfied in giant planetary atmospheres across all temperatures, the analytic approach allows one to derive a representative atmosphere-averaged scattering albedo from the observables, and has been successfully used in fitting observations of solar-system planets (Kattawar \& Adams 1971; Sromovsky 2005; Schmid et al. 2006).

Many researchers have derived analytic\footnote{By `analytic' we mean solutions that can be expressed in closed form or  which can be evaluated to arbitrary precision. This includes results derived using ordinary (as opposed to partial) differential equations.} expressions for reflection coefficients, albedos, phase curves, and polarization parameters for different forms of scattering in planetary atmospheres (Chandrasekhar 1950; Horak 1950; Abhayankar \& Fymat 1970; Kattawar \& Adams 1971;  Van de Hulst 1981; Bhatia \& Abhayankar 1982). Chandrasekhar (1950) and Horak (1950) reported analytic solutions for reflected specific intensities from homogeneous atmospheres for the cases of isotropic scattering, asymmetric scattering, and conservative Rayleigh scattering ($\omega = 1.0$), all for finite and semi-infinite atmospheres. Chandrasekhar (1960) and Horak \& Chandrasekhar (1960) further generalized their solutions to address non-conservative Rayleigh scattering ($\omega < 1$) in semi-infinite atmospheres. Formulations using the scalar phase function for single-scattering do not take into account the effect of polarization. Chandrasekhar (1960) and Abhayankar \& Fymat (1970,1971) reported a general solution including conservative and non-conservative Rayleigh scattering in homogeneous semi-infinite atmospheres using the full Rayleigh phase matrix, so that the Stokes parameters (Stokes 1852; Chandrasekhar 1950,1960), and hence the polarization, could also be computed analytically. Subsequent studies have applied analytic approaches to a wide range of problems in planetary and exoplanetary atmospheres (Kattawar \& Adams 1974;  Van de Hulst 1980; Bhatia \& Abhayankar 1982; Sromovsky 2005; Schmid et al. 2006; Natraj et al. 2009; Kane \& Gelino 2010). 

In this work, we present a systematic procedure to compute observables of reflected light for the different scattering phenomena under some basic assumptions. We consolidate the formalism, solution techniques, and results from analytic models that are available in the literature, but are often in scattered and obscure forms. Building on the tradition of Chandrasekhar (1950,1960), we follow the H-function approach to represent emergent fluxes for different scattering phenomena. We consider cloud-free homogeneous semi-infinite atmospheres, scattering in accordance with different scattering phase functions. We consider both conservative ($\omega = 1.0$) and non-conservative ($\omega <  1.0$) scattering. We compute geometric albedos and phase curves for Rayleigh scattering, with scalar and vectorial phase functions, isotropic scattering, asymmetric scattering, and Lambert scattering. We also compute polarization curves for the case of Rayleigh scattering. 

We provide a step-by-step procedure to obtain the disk-integrated emergent fluxes and polarization, with the appropriate angular transformations to the celestial reference frame. The phase curves and polarization curves are also computed as a function of the mean anomaly, or a time coordinate, for given orbital parameters. We also provide analytic fits to the geometric and spherical albedos for Rayleigh scattering as a function of the scattering albedo. Thus, given an observed geometric albedo at a certain wavelength, an average scattering albedo for the atmosphere can be estimated, which can then be used to predict other observables such as the phase curve and the Stokes parameters. 

In what follows, we first describe the formalism in Section~\ref{sec:methods}. We present our results on 
phase curves and albedos for different scattering phase functions in Section~\ref{sec:results}. In Section~\ref{sec:results_inclination}, we study the dependence of polarization curves for Rayleigh scattering atmospheres 
on the orbital parameters. We summarize our results in Section~\ref{sec:summary}. 

\begin{figure}[ht]
\centering
\includegraphics[width = 0.5\textwidth]{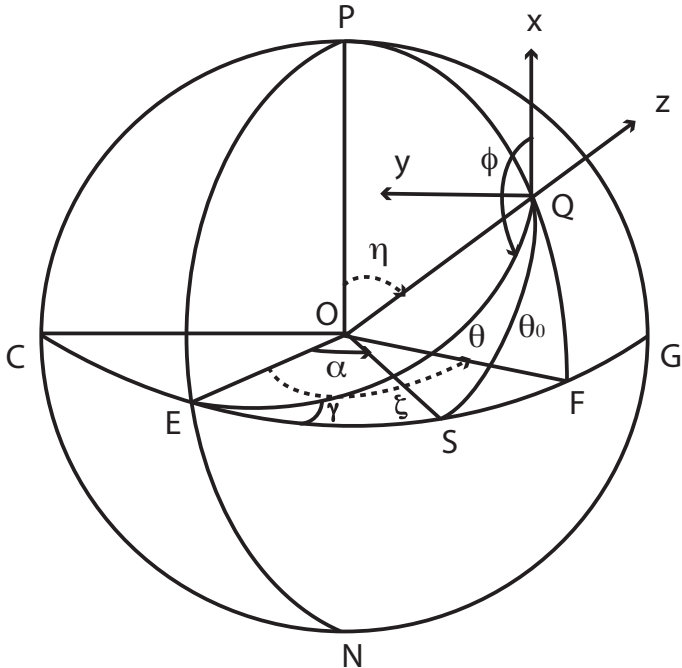}
\caption{Planetary coordinate system. See Section~\ref{sec:geometry} for description.} 
\label{fig:coords}
\end{figure}

\section{Methods}
\label{sec:methods}
Our goal in this study is to provide all the components of an analytic framework with which to interpret 
observations of reflected light from extrasolar planets. In this section, we briefly review some basic 
definitions of observables and describe the models used in this work. 

\subsection{Definitions}

The spherical albedo ($A_s$) of a planet, which is the total fraction of incident light reflected by a sphere at 
all angles, for an incident flux $\pi F$, is given by:
\begin{equation}
A_s = 2 \int^\pi_0 \dfrac{j(\alpha)}{\pi F} \sin\alpha~d\alpha, 
\end{equation}
where $j(\alpha)$ is the emergent flux and $\alpha$ is the phase angle, defined as the angle between the 
star, planet, and observer, whose vertex lies at the planet. 
$A_s$ can also be expressed in terms of two other quantities of interest, the geometric albedo ($A_g$) and 
the phase integral ($q$), given by:
\begin{eqnarray}
A_g = \dfrac{j(0)}{\pi F} \textrm{~~and~~} q = 2 \int^\pi_0 \dfrac{j(\alpha)}{j(0)} \sin\alpha~d\alpha.
\label{eq:Ag}
\end{eqnarray}

The quantity $\Phi (\alpha) = j(\alpha)/j(0)$ is the  classical phase function\footnote{The phase function, $\Phi (\alpha)$, should not be confused with the scattering phase function, $p(\cos\Theta)$ (see e.g. Section~\ref{sec:iso} and ~\ref{sec:asymm}).}. The planet-star flux ratio 
observed at Earth as a function of the phase angle is given by 
\begin{equation}
\frac{F_p}{F_\star} = A_g \left(\frac{R_p}{a}\right)^2 \Phi (\alpha), 
\label{eq:fluxratio} 
\end{equation}
where $R_p$ is the planetary radius and $a$ is the orbital separation. 

\subsection{Orbital Phase}
\label{sec:orb_phase}
Given a planetary phase curve as a function of the phase angle, the observed phase curve 
as a function of the orbital phase can be computed using Kepler's law.  
The phase angle ($\alpha$) is related to the true anomaly ($\theta$) and the orbital inclination ($i$) by 
\begin{equation}
\cos\alpha = \sin(\theta+\omega_p) \sin i, 
\end{equation}
where $\omega_p$ is the argument of periastron. 
The true anomaly is related to the orbital phase via Kepler's law: 
\begin{equation}
M = E - e \sin E.
\label{eq:kepler}
\end{equation}
Here, $M$ is the mean anomaly, given by $2\pi(t-t_p)/P$; $t$ is time; $t_p$ is time of pericenter passage;  
and $P$ is the orbital period. $e$ is the orbital eccentricity and $E$ is the eccentric anomaly. $E$ can be 
expressed in terms of the true anomaly as: 
\begin{equation}
\sin E = \frac{\sin\theta \sqrt{1-e^2}}{1+e\cos\theta}. 
\end{equation}

The Kepler 
equation can be rewritten as: 
\begin{equation}
t = \frac{P}{2\pi}\Bigg[2 \tan^{-1} \Bigg(\sqrt{\frac{(1-e)}{(1+e)}}\tan\frac{\theta}{2}\Bigg) - \frac{e\sin\theta\sqrt{1-e^2}}{1+e\cos\theta}\Bigg].
\end{equation}
Here, $t = 0$ corresponds to pericenter passage. Thus, the planetary phase curve $\Phi(\alpha)$ can be 
expressed as $\Phi(t)$. 

\subsection{Emergent Intensity}

The emergent intensity, $j (\alpha)$, from a planetary atmosphere as a function of the phase angle determines 
all the reflected light observables, as is evident from Eqs.~(\ref{eq:Ag}) \& (\ref{eq:fluxratio} ). Theoretical computation of $j(\alpha)$ involves solving the radiative transfer equation while including the required sources of scattering and absorption in the atmosphere. Several numerical codes exist that solve the general radiative transfer problem of scattering in inhomogeneous atmospheres, i.e. where the scattering albedo ($\omega$) is a function of pressure ($P$) and temperature ($T$) in the atmosphere, to compute albedo spectra and phase curves (Marley et al. 1999; Seager et al. 2000; Sudarsky et al. 2000, 2005; Burrows et al. 2008). The sources of scattering typically include Rayleigh scattering, Mie scattering due to condensates if present, along with a weaker contribution due to Raman scattering.  For cloud-free atmospheres, as are considered likely for highly-irradiated giant planets (Seager \& Sasselov 1998; Sudarsky et al. 2000), Rayleigh scattering dominates the scattering. In this work, we consider homogeneous (uniform $\omega$) and semi-infinite atmospheres for which analytic expressions can be derived for $j(\alpha)$ for several scattering conditions (Russell 1916; Chandrasekhar 1950; Horak 1950; Kattawar \& Adams 1971), with particular emphasis on vector Rayleigh scattering. 

\subsubsection{Geometry}
\label{sec:geometry}

The planetary coordinate system is shown in Fig.\ref{fig:coords}. O is the center of the planet and OE and OS 
denote the directions from the planet to the Earth and the star, respectively. The angle EOS is the phase angle ($\alpha$). 
CESG defines the planetary equator, and P and N are the poles. The longitude ($\zeta$) and latitude ($\eta$) on the 
planet sphere, shown as dotted lines in Fig.\ref{fig:coords}, are defined with respect to the great circles PEN and CESG, respectively. Q is an arbitrary point on the surface of the planet, shown for illustration, where a local coordinate system can be defined as shown. In this system, the $z$-axis is normal to the surface at Q. The incident ray from the star at the point Q is defined by the angles of incidence ($\theta_0$, $\phi_0$) with respect to the local coordinate system. The $x$-axis is chosen such that $\phi_0 = 0$, and the $y$-axis follows from the $x$ and $y$ axes assuming a right-handed Cartesian coordinate system. The direction of the reflected ray detected at Earth is defined by ($\theta$, $\phi$), as shown in Fig.\ref{fig:coords}. The angles of the incident and reflected rays with respect to the normal, $\theta_0$ and $\theta$, are typically represented in terms of their direction cosines $\mu_0 = \cos\theta_0$ and $\mu = \cos\theta$. Vectorial quantities in the local frame at Q can be transformed to those in a global coordinate system centered at O by a rotation by an angle $\gamma$, as discussed in section~\ref{sec:transform} below. 

\subsubsection{Integration over the Visible Crescent}

At each phase angle in the orbit, in order to compute the total emergent intensity $j(\alpha)$ from the planetary disk, we need to integrate the specific intensity over the illuminated surface of the disk. In planetary coordinates, we 
denote the specific intensity 
of emergent radiation by $I(\eta,\zeta)$, where $\eta$ and $\zeta$ are the planetary latitude and longitude, 
respectively, following the notation of Horak (1950) and Kattawar \& Adams (1971). If $I(\eta,\zeta)$ is known at 
each location on the disk, the disk-integrated emergent intensity at a given phase angle ($\alpha$) is given  
by: 
\begin{equation}
j(\alpha) = \int_0^\pi d\eta~\sin^2\eta \int_{\alpha-\pi/2}^\pi d\zeta~I(\eta,\zeta)~\cos\zeta.
\label{eq:flux1}
\end{equation}

The planetary coordinates ($\eta$,~$\zeta$) are related to the angles of incidence and reflection by 
\begin{eqnarray}
\mu_0 &=& \sin\eta~\cos(\zeta - \alpha) \\ 
\mu &=& \sin\eta~\cos\zeta.
\end{eqnarray}

The integral in Eq.(\ref{eq:flux1}) can be evaluated numerically by performing a coordinate transformation of Eq. (\ref{eq:flux1}) to obtain the following equation: 

\begin{equation}
j(\alpha) = \dfrac{(\cos\alpha + 1)}{2} \int_{-1}^{+1} d\psi \sqrt{1 - \psi^2}\int_{-1}^{+1} d\xi~I(\psi,\xi),
\label{eq:flux2}
\end{equation}

where, $\psi = \cos \eta$, and 
\begin{equation}
\xi = \left(\dfrac{2}{\cos\alpha + 1}\right)\nu +  \left(\dfrac{\cos\alpha - 1}{\cos\alpha + 1}\right), 
\end{equation}
where $\nu = \sin\zeta$. 

The double integral in Eq.(\ref{eq:flux2}) can then be evaluated using standard quadrature methods as:
\begin{equation}
j(\alpha) = \dfrac{(\cos\alpha + 1)}{2} \sum_{i=1}^n \sum_{j=1}^n w_i u_j I(\psi_i,\xi_j), 
\label{eq:j_alpha}
\end{equation}
where, $w_i$ and $u_j$ are the quadrature weights for the corresponding abscissae $\psi_i$ and $\xi_j$, respectively  (Horak 1950; Kattawar \& Adams 1971). In the present work, we use a 32-point Gaussian quadrature for the integral over $\xi$, and a 32-point quadrature using Chebyshev polynomials for the $\psi$ integral. While the abscissae and weights for the Gaussian quadrature are available in standard tables, the same for the Chebyshev polynomials can be evaluated simply by: $\psi_i = \cos[i\pi/(n+1)], ~~ w_i = [\pi/(n+1)]\sin^2[i\pi/(n+1)]$.

\subsection{Specific Intensity for Different Scattering Phenomena}
\label{sec:specific_intensity}
Before computing the total emergent intensity from the illuminated surface of the planetary 
disk, we need to be able to compute the emergent specific intensity at any arbitrary point on 
the planetary surface. Given an incident ray with coordinates ($\theta_0$, $\phi_0$) at a point 
on the planetary surface, as described in Section~\ref{sec:geometry}, our goal is to compute the 
reflected intensity in the direction ($\theta$, $\phi$). We shall refer to $\theta$ and $\theta_0$ by 
their direction cosines $\mu$ and $\mu_0$.  

\subsubsection{Lambert Scattering}

One of the simplest models usually considered is that of an isotropically scattering Lambertian surface. 
The emergent specific intensity for such a surface (Chandrasekhar 1960) is given by: 
\begin{equation}
I(\mu) = \omega F \mu, 
\label{eq:lambert_I}
\end{equation} 
where (as defined earlier) $\omega$ is the scattering albedo and $\pi F$ is the incident stellar flux. 
As is evident from Eq.(\ref{eq:lambert_I}), the reflected intensity is independent of the angle of incidence, 
and depends solely on the direction cosine ($\mu$) of the observer. 

\subsubsection{Isotropic Scattering}
\label{sec:iso}

Let us consider the case of a semi-infinite atmosphere scattering isotropically with a 
scattering phase function given by $p(\cos\Theta) = \omega$. For an incident flux 
($\pi F$) and a given angle of incidence ($\mu_0 = \cos \theta_0$), the emergent specific intensity 
at the top of the atmosphere (at optical depth $\tau = 0$) along a given direction ($\mu = \cos \theta$) 
is given by (Chandrasekhar 1950, 1960; Horak 1950)\footnote{We replace the notation $I(0; \mu)$ of 
previous works (Chandrasekhar 1950, 1960; Horak 1950; Abhyankar \& Fymat 1970) with $I(\mu)$, 
since we are concerned only with the emergent intensity for which $\tau = 0$ is implied.}:
\begin{equation}
I(\mu) = \frac{\omega F}{4} \frac{\mu_0}{\mu_0 + \mu}H(\mu) H(\mu_0). 
\label{eq:iso}
\end{equation} 
The function $H (\mu)$ satisfies the integral equation:
\begin{equation}	
	\dfrac{1}{H(\mu)} =	\left[ 1 - 2 \int_0^1 \Psi(\mu) d\mu\right]^{\frac{1}{2}}+ \int_0^1 H(\mu^\prime) \Psi(\mu^\prime) \frac{\mu^\prime d\mu^\prime}{\mu + \mu^\prime},
\label{eq:eq_H}
\end{equation}
which can be solved by iteration until convergence is achieved. We found a reasonable initial condition to be $H(\mu) = 1$, and the integrals in the equation can be evaluated numerically by standard quadrature methods; we used a 32-point Gaussian quadrature. $\Psi(\mu)$ is known as the characteristic function, and is given 
by $\Psi(\mu) = \omega = $ constant, for the case of isotropic scattering. Thus, the emergent intensity 
at a given point on the planetary 
sphere can be computed for any angles of incidence and reflection. In order to compute the intensity 
detected by an observer at Earth, the emergent intensity in the direction of the observer is integrated 
over the illuminated crescent of the planet, as described in Section~\ref{sec:geometry}

\subsubsection{Asymmetric Scattering}
\label{sec:asymm}. 

The scattering phase function for asymmetric scattering is given by $p(\cos\Theta) = \omega (1 + x\cos\Theta)$, where $\Theta$ is 
the angle between the incident and scattered rays, and $x$ is the asymmetry factor, $x \in [0,1]$. In analogy with Eq.(\ref{eq:iso}), the emergent specific intensity is given by (Chandrasekhar 1960): 
\begin{multline}
I (\mu,\phi)  = \frac{\omega F}{4} \frac{\mu_0}{\mu_0 + \mu}\Big[\psi(\mu)\psi(\mu_0) - x \Phi(\mu)\Phi(\mu_0)  \\ 
+ x\sqrt{(1-\mu^2)(1-\mu_0^2)} H^{(1)}(\mu) H^{(1)}(\mu_0)\cos(\phi-\phi_0)\Big], 
\label{eq:asymm}
\end{multline}
where, $\psi(\mu) = H(\mu)(1-c\mu)$ and $\Phi(\mu) = q\mu H(\mu)$, and the characteristic functions with which to derive $H(\mu)$ and 
$H^{(1)}(\mu)$ are $\frac{1}{2}\omega[1+x(1-\omega)\mu^2]$ and $\frac{1}{4}x\omega(1-\mu^2)$, respectively. $q$ and $c$ are given by $q = 2(1-\omega)/(2 - \omega\alpha_0)$ and $c = x\omega(1-\omega)\alpha_1/(2 - \omega\alpha_0)$, where $\alpha_0$ and $\alpha_1$ are the zeroth and first integral moments of $H(\mu)$: $\alpha_0 = \int_0^1 H(\mu) d\mu$ and $\alpha_1 = \int_0^1 \mu H(\mu) d\mu$.

\subsubsection{Rayleigh Scattering - Scalar formalism}

Rayleigh scattering constitutes a dominant scattering mechanism in planetary atmospheres, 
especially in the absence of Mie scattering by clouds. The scattering phase function for Rayleigh 
scattering is given by $p(\cos\Theta) = \frac{3}{4}\omega (1 + \cos^2\Theta)$. This scalar form of the phase 
function does not account for the changes in polarization in the reflected beam during multiple scatterings in 
the atmosphere. Nevertheless, it is customary in the literature to adopt the scalar phase function for Rayleigh scattering for ease of computation (Marley et al. 1999; Sudarsky et al. 2000, 2005; Burrows et al. 2008). We discuss the emergent specific intensity with the scalar Rayleigh function in this 
section, and we discuss the full vectorial treatment in Section~\ref{sec:rayleigh_vector}. 

Horak \& Chandrasekhar (1961) solved the case of a semi-infinite homogeneous atmosphere 
with a single-scattering scalar phase function given by 
\begin{equation}
p(\cos\Theta) = \omega + \omega_1 P_1(\cos\Theta) + \omega_2 P_2(\cos\Theta), 
\label{eq:rayleigh_scalar_full}
\end{equation}
where, the functions $P_1(x)$ and $P_2(x)$ are Legendre Polynomials of order 1 and 2, respectively, and 
$\omega$, $\omega_1$, and $\omega_2$ are constants. 
The emergent specific intensity for the case of Rayleigh-like scattering in a semi-infinite atmosphere can be obtained 
from their results by setting $\omega_1 = 0$ and $\omega_2 = \omega/2$ in Eq. (\ref{eq:rayleigh_scalar_full}) to 
obtain the scalar phase function 
\begin{equation}
p(\cos\Theta) = \frac{3}{4}\omega(1 + \cos^2\Theta). 
\label{eq:rayleigh_scalar}
\end{equation}

We note that the closed form expressions for the specific intensity summarized in Section~VIII of Horak \& Chandrasekhar (1961) have some errors. Therefore, we choose to re-derive the specific intensity and obtained the following form for the solution, similar to Eqs. (\ref{eq:iso}) and (\ref{eq:asymm}):  
\begin{multline}
I (\mu, \phi)  = \frac{\omega F}{4} \frac{\mu_0}{\mu_0 + \mu}\Big[\Phi(\mu)\Phi(\mu_0) + \frac{1}{8}\psi(\mu)\psi(\mu_0) \\
- \frac{3}{2} \mu\mu_0 \sqrt{(1-\mu^2)(1-\mu_0^2)} H^{(1)}(\mu) H^{(1)}(\mu_0)\cos(\phi-\phi_0) \\ 
+ \frac{3}{8}(1-\mu^2)(1-\mu_0^2)H^{(2)}(\mu) H^{(2)}(\mu_0) \cos2(\phi-\phi_0)\Big], 
\label{eq:scalar}
\end{multline}
where, $\Phi(\mu) = \mu H^{(0)}(\mu) \Phi_1(\mu)$ and $\psi(\mu) = H^{(0)}(\mu)[3+\psi_1(\mu)]$. $H^{(0)}(\mu)$, $H^{(1)}(\mu)$ and $H^{(2)}(\mu)$ are H-functions with characteristic functions given by $\frac{3}{16}\omega[3 - (2-\omega)\mu^2 + 3(1-\omega)\mu^4]$, $\frac{3}{8}\omega(\mu^2 - \mu^4)$, and $\frac{3}{32}\omega(1-\mu^2)^2$, respectively. $\Phi_1(\mu)$ and $\psi_1(\mu)$ satisfy the coupled integral equations:
\begin{multline}
\Phi_1(\mu) = \frac{\mu}{H^{(0)}(\mu)} + \frac{\omega}{2}\int_0^1\frac{\mu^2H(\mu^\prime)}{\mu+\mu^\prime}
\Big[\mu\mu^\prime\Phi_1(\mu)\Phi_1(\mu^\prime) \\ + \frac{1}{8}(3+\psi_1(\mu))(3+\psi_1(\mu^\prime))\Big]d\mu^\prime \label{eq:s1}
\end{multline}
and
\begin{multline}
\psi_1(\mu) + 3 = \frac{3-\mu^2}{H^{(0)}(\mu)} + \frac{\omega}{2}\int_0^1\frac{\mu H(\mu^\prime)}{\mu+\mu^\prime}
\Big[\mu\mu^\prime\Phi_1(\mu)\Phi_1(\mu^\prime) \\ + \frac{1}{8}(3+\psi_1(\mu))(3+\psi_1(\mu^\prime))\Big] (3-{\mu^\prime}^2)  d\mu^\prime. 
\label{eq:s2}
\end{multline}
The integral equations (\ref{eq:s1}) and (\ref{eq:s2}) can be solved simultaneously, for $\Phi_1(\mu)$ and $\psi_1(\mu)$, by iteration until convergence is achieved. We used 32-point Gaussian quadrature for the integrals. 

\subsubsection{Rayleigh Scattering - Vector formalism}
\label{sec:rayleigh_vector}
In this section, we describe in detail the formalism with which to obtain the intensities of reflected radiation 
in directions perpendicular and parallel to 
the plane of incidence of a homogeneous and semi-infinite Rayleigh scattering atmosphere with arbitrary scattering 
albedo ($\omega$).  We follow the methodology of Abhyankar and Fymat (1970), who used the full Rayleigh phase matrix to develop an analytic formalism for scattering based on the H-function approach of Chandrasekhar (1950). 

The different components of reflected intensity, which are essentially elements of a modified Stokes vector (Stokes 1852; Chandrasekhar 1950,1960), are given by:
\begin{eqnarray}
I_{l} (\mu, \phi) &=& \omega D \left[A_l + B_l \cos(\phi_0 - \phi) + C_l\cos[2(\phi_0 - \phi)\right] \mu_0F \label{eq:I_nonconsv_1} \nonumber \\
I_{r} (\mu, \phi) &=& \omega D \left[A_r + C_r\cos[2(\phi_0 - \phi)\right] \mu_0F \\
U (\mu, \phi) &=& \omega D \left[B_U \sin(\phi_0 - \phi) + C_U\sin[2(\phi_0 - \phi)\right] \mu_0F \nonumber\\
\textrm{and}&&V(\mu, \phi) = 0~. \nonumber
\end{eqnarray}

The Stokes vector is a four-element vector which describes the intensity and polarization of a beam of light, and is represented as ${\textbf I} = \left[ I~Q~U~V \right]$. Here, $I$ is the total intensity, and $Q$ and $U$ are 
components of intensities with linear polarization, and $V$ is the intensity with circular polarization. In terms of  
the modified Stokes parameters in Eq.(\ref{eq:I_nonconsv_1}), $I = I_l + I_r$ and $Q = I_l - I_r$, where $I_l$ and $I_r$ are 
the components of intensity parallel and perpendicular to the plane of incidence. The different terms in Eq.(\ref{eq:I_nonconsv_1}) are given by: 
\begin{eqnarray}
&&A_l = N_1(\mu)[N_1(\mu_0) + N_3(\mu_0)] + N_2(\mu)[N_2(\mu_0) + N_4(\mu_0)] \nonumber\\
&&A_r = N_3(\mu)[N_1(\mu_0) + N_3(\mu_0)] + N_4(\mu)[N_2(\mu_0) + N_4(\mu_0)] \nonumber \\
&&B_U = 3 \mu_0(1 - \mu^2)^{1/2}(1 - \mu_0^2)^{1/2}H^{(1)}(\mu)H^{(1)}(\mu_0) \nonumber \\
&&B_l = -\mu B_U \nonumber \\
&&C_r = \dfrac{3}{4}(1 - \mu_0^2)H^{(2)}(\mu)H^{(2)}(\mu_0) \label{eq:I_nonconsv_2}
\\
&&C_l = -\mu^2 C_r \nonumber \\
&&C_U = 2\mu C_r  \nonumber\\
&&\textrm{and } \nonumber \\
&&D = 1/[8(\mu + \mu_0)] \textrm{    .} \nonumber
\end{eqnarray}

The functions $N_i(\mu)$ and $H^{(i)}(\mu)$ are computed along the lines of the general H-function 
approach of Chandrasekhar (1950,1960). The $N_i(\mu)$ satisfy the matrix equation:
\begin{equation}	
	{\bf N}(\mu) = {\bf M}(\mu) + \frac{1}{2} \omega \mu {\bf N}(\mu) \int_0^1 {\bf N}^T(\mu^\prime) {\bf M}(\mu^\prime) \frac{d\mu^\prime}{\mu + \mu^\prime},	
\label{eq:eq_N}
\end{equation}
where, ${\bf N} = \left[ \begin{array}{cc} N_1 & N_2 \\ N_3 & N_4 \end{array} \right]$, ${\bf N}^T$ is the transpose of ${\bf N}$, \newline 
and
\begin{equation}
{\bf M} = \left[ \begin{array}{cc} M_1 & M_2 \\ M_3 & M_4 \end{array} \right] = \dfrac{\sqrt{3}}{2} 
\left[ \begin{array}{cc} \mu^2 & \sqrt{2}(1-\mu^2) \\ 1 & 0 \end{array} \right].
\end{equation} 

Eq.(\ref{eq:eq_N}) can be solved numerically by iteration until convergence is achieved. A 
reasonable initial condition is ${\bf N} = {\bf M}$. The integration on the right-hand-side of 
Eq.(\ref{eq:eq_N}) is easily evaluated using a 32-point Gaussian quadrature. 

The H-functions $H^{(1)}$ and $H^{(2)}$ are solutions of Eq. (\ref{eq:eq_H}) for characteristic functions:
\begin{equation}
\Psi^{(1)}(\mu) = \dfrac{3}{8} \omega (1-\mu^2)(1+2\mu^2)
\end{equation}
and 
\begin{equation}
\Psi^{(2)}(\mu) = \dfrac{3}{16} \omega (1+\mu^2)^2.
\end{equation}

\subsubsection{Transformation from local coordinates to celestial coordinates}
\label{sec:transform}

The intensities in Section~\ref{sec:rayleigh_vector} are defined with respect to a differential area element in the 
local meridional plane, i.e. the plane containing the incident and reflected rays at a local patch on the disk. These 
intensities need to be transformed into the global plane of reference of the planet as a whole with respect to the 
observer. The transformation is obtained by rotating the Stokes vector in the local plane through an angle 
$\gamma$ given by: 
\begin{equation}
\cos\gamma = \sin\eta~\sin\zeta/\sin\theta \textrm{, ~or~} \sin\gamma = \cos\eta/\sin\theta.
\end{equation}

The rotation matrix for transforming \textbf{I} to the celestial frame is given by: 
\begin{equation}
{\textbf L} = \left[\begin{array}{cccc} 1 & 0 & 0 & 0 \\ 0 & \cos 2\gamma & \sin 2\gamma & 0 \\ 
0 & -\sin 2\gamma & \cos 2\gamma & 0 \\ 0 & 0 & 0 & 1 \end{array} \right].
\label{eq:stokes_transform} 
\end{equation}

The transformed intensities are given by ${\textbf I^\prime = \textbf{L}\cdot\textbf{I}}$, with the result 
that: 
\begin{eqnarray}
I^\prime &=& I \nonumber\\
Q^\prime &=& Q \cos 2\gamma + U \sin 2\gamma \label{eq:transform}\\
U^\prime &=& - Q \sin 2\gamma + U \cos 2\gamma \nonumber\\
V^\prime &=& V~. \nonumber
\end{eqnarray}
Since, $I^\prime = I_l^\prime + I_r^\prime$ and $Q^\prime = I_l^\prime - I_r^\prime$ by definition, the transformed $I_l^\prime$ and 
$I_r^\prime$ are given by: 
\begin{equation}
I_l^\prime = (I^\prime + Q^\prime)/2 \textrm{ and } I_r^\prime = (I^\prime - Q^\prime)/2.  
\label{eq:transform_1}
\end{equation} 

These new intensities of the rotated Stokes vector are then used in Eq.(\ref{eq:flux2}) to compute the disk-integrated flux in the planetary frame. 

The degree of polarization ($P$) is defined as $P = \sqrt{Q^2 + U^2}/I$. For edge-on orbits, however, the 
disk-integrated Stokes U vanishes due to symmetry in the north-south direction, in which case $P = Q/I$. 
The angle of polarization is defined as $\theta_P = \frac{1}{2}\tan^{-1}(U/Q)$.

\begin{figure}[h]
\centering
\includegraphics[width = 0.5\textwidth]{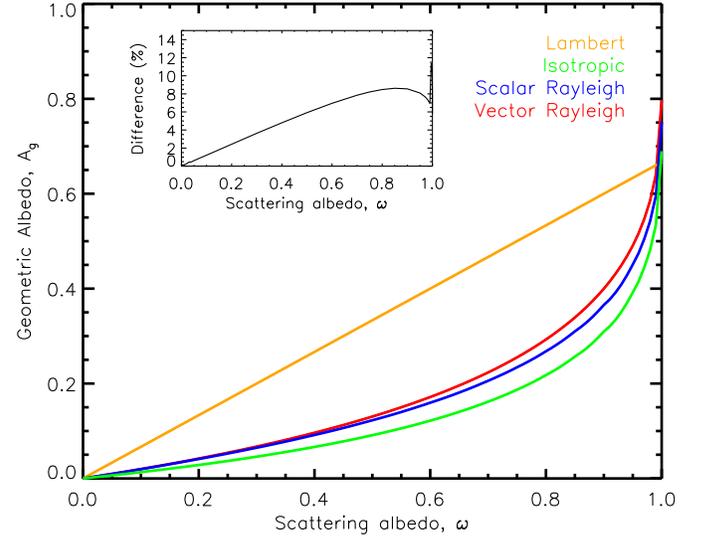}
\caption{Geometric albedos as a function of scattering albedo for different scattering phase functions. The red and blue curves in the main panel show geometric albedos for Rayleigh scattering using the full phase matrix and using only the scalar phase function, respectively. The inset shows the percent difference between the two curves. The green and orange curves in the main panel correspond to isotropic and Lambert scattering, respectively.} 
\label{fig:Ag}
\end{figure}

\begin{figure}[h]
\centering
\includegraphics[angle=-90,width = 0.5\textwidth]{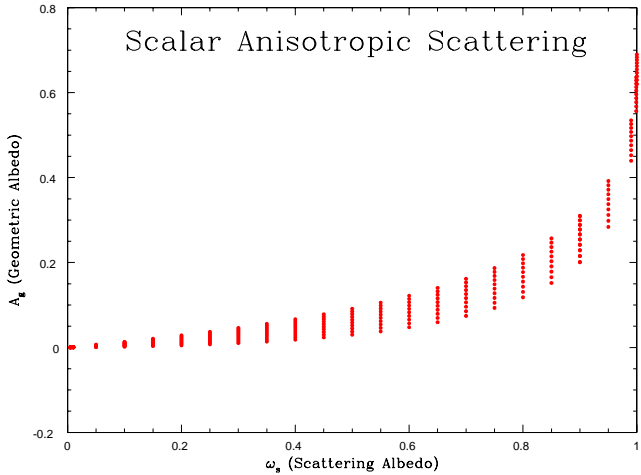}
\caption{Geometric albedos as a function of scattering albedo for the asymmetric scattering phase functions. For 
each $\omega$, geometric albedos are shown for a range of asymmetry factors, $x$, between 0 and 1.} 
\label{fig:Ag_asymm}
\end{figure}

\section{Results}
\label{sec:results}

In this section, we present geometric albedos and phase curves, and, for the vector Rayleigh case, 
polarization curves, of reflected light for different scattering phenomena. As discussed in Section~\ref{sec:specific_intensity}, 
we consider the following cases: (a) Lambert scattering, (b) isotropic scattering, (c) asymmetric scattering, 
(d) Rayleigh scattering with scalar phase function, and (e) Rayleigh scattering using the phase matrix, in which 
case the polarization can be computed. The geometric albedo ($A_g$) and phase curve, $\Phi(\alpha)$, for each 
case are obtained by the following steps: 

\noindent
\begin{enumerate}
\item{For a given $\alpha$, compute disk integrated emergent flux, $j(\alpha)$, using Eq.(\ref{eq:j_alpha}):
\begin{list}{\labelitemi}{\leftmargin=1em}
\item{Loop over summations: $i=1\rightarrow m$, $j=1\rightarrow n$, in Eq.~(\ref{eq:j_alpha}).}
\item{Transform each ($\psi_i$,~$\xi_j$) to the corresponding ($\mu$,~$\phi$)}
\item{Compute intensity $I(\psi_i,~\xi_j)$ $\equiv$ $I(\mu,~\phi)$ for the given scattering case, using Eqs. (\ref{eq:iso}), (\ref{eq:asymm}), (\ref{eq:scalar}), or (\ref{eq:I_nonconsv_1}).}
\item{In the case of Rayleigh scattering with the vector formulation: 
\begin{itemize}
\renewcommand{\labelitemi}{$-$}
\item{Compute the different intensities of the Stokes's vector in the planet frame [$I_r(\mu,~\phi)$, 
          $I_l(\mu,~\phi)$, $U(\mu,~\phi)$, $V(\mu,~\phi)$] using Eqs. (\ref{eq:I_nonconsv_1}) and (\ref{eq:I_nonconsv_2}).} 
\item{Transform the intensities to the celestial frame using Eqs. (\ref{eq:transform}) and (\ref{eq:transform_1}).}
\item{Use the transformed intensities in computing different components of $j(\alpha)$.}
\end{itemize}
\item{Compute the final sum with the appropriate weights as shown in Eq.(\ref{eq:j_alpha}).}
}
\end{list}
}
\item{Repeat Step 1 for $\alpha = 0 \rightarrow \pi$ and derive $A_g$, $\Phi(\alpha)$, and polarization.
\begin{list}{\labelitemi}{\leftmargin=1em}
\item{Compute $A_g = j(0)/\pi F$, from Eq. (\ref{eq:Ag})}
\item{Compute the phase curve, $\Phi(\alpha) = j(\alpha)/j(0)$.}
\item{For vector Rayleigh, compute polarization (see Section~\ref{sec:transform}).}
\end{list}
}
\item{To obtain phase curves as a function of time ($t$), compute $\alpha$ for given $t$ (see Section~\ref{sec:orb_phase}.})
\end{enumerate}

\subsection{Lambert Scattering}

The emergent flux for Lambert scattering is fully analytic (Russell 1916; Chandrasekhar 1950,1960) and 
is given by:
\begin{equation}
j(\alpha) = \frac{2}{3} \omega \pi F \big[\frac{\sin(\alpha) + (\pi - \alpha)\cos(\alpha)}{\pi}\big] ,
\label{eq:lambert}
\end{equation}
where $\omega$ is the scattering albedo and $\pi F$ is the incident stellar flux. 
Thus, the geometric albedo and phase function of a Lambertian surface are given by  
$A_g = \frac{2}{3}\omega$, and $\Phi(\alpha) = [\sin(\alpha) + (\pi - \alpha)\cos(\alpha)]/\pi$. For a perfectly 
reflecting Lambertian surface (i.e. $\omega = 1.0$) the familiar $A_g = 2/3$ is obtained, and the 
spherical albedo is given by $A_s = 1$. As shown in Fig.~\ref{fig:Ag}, the Lambert phase function 
leads to higher $A_g$ for all $\omega$, compared to all the other scattering mechanisms considered 
in this work. The phase curve for Lambert scattering is shown in Figure~\ref{fig:pha_all}. 

\subsection{Istrotropic and Asymmetric Scattering}

The cases of isotropic and asymmetric scattering correspond to scattering phase functions that 
are linear in $\omega$, as discussed in Sections~\ref{sec:iso} and \ref{sec:asymm}. The isotropic phase function 
is a special case of the asymmetric phase function, with an asymmetry factor ($x$) of 0. The geometric albedo 
as a function of the scattering albedo for scattering in accordance with the isotropic and asymmetric phase 
functions are shown in Figs.~\ref{fig:Ag} and \ref{fig:Ag_asymm}, respectively.  As shown in Fig.~\ref{fig:Ag}, 
isotropic scattering leads to lower $A_g$ compared to that due to the Lambert and Rayleigh phase functions, 
for all $\omega$. Figure~\ref{fig:pha_all} shows the phase curves for isotropic scattering for different $\omega$.
On the other hand, as shown in Fig.~\ref{fig:Ag_asymm}, asymmetric scattering with $x > 0$ leads to 
higher values of $A_g$, compared to isotropic scattering, which for $x \sim 1$ are comparable to those 
obtained from the Rayleigh phase function. 

\subsection{Scalar and Vector Rayleigh scattering}
\label{sec:results_rayleigh}
\begin{figure}[ht]
\centering
\includegraphics[width = 0.5\textwidth]{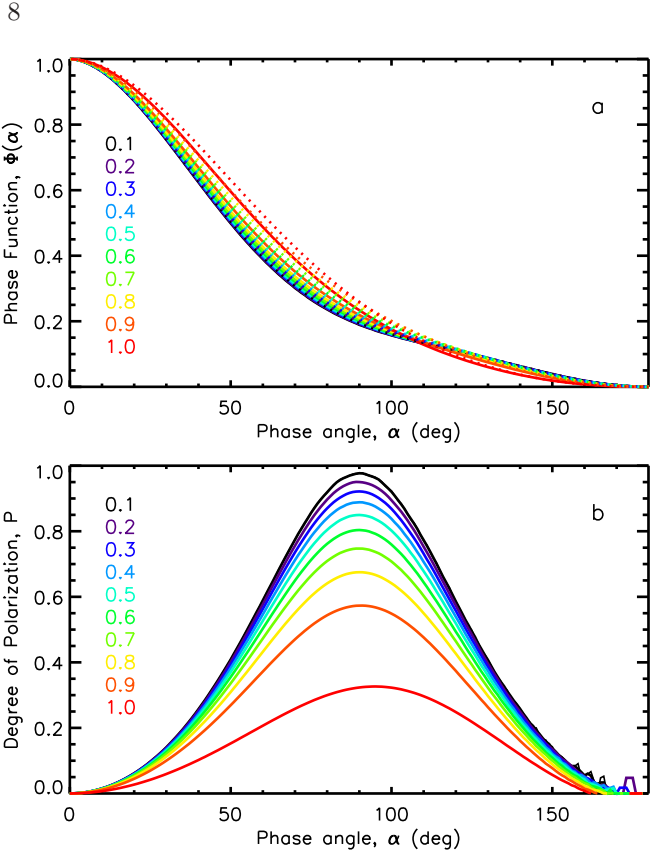}
\caption{Phase curves and polarization for Rayleigh scattering. The 
different curves in each panel correspond to the different scattering albedos shown in the legend. In the 
upper panel, the solid curves show the phase curves for Rayleigh scattering using the vectorial Rayleigh phase matrix, 
whereas the dotted curves show those with the scalar phase function which does not incorporate polarization. 
The lower panel shows the degree of polarization ($P$), using the Rayleigh phase matrix. $P$ is defined as 
$P = \sqrt{Q^2 + U^2}/I$, where $Q$ and $U$ are the two Stokes parameters for linear polarization and $I$ is 
the total intensity. For all the curves shown here, the orbit is assumed to be edge on ($i = 90^\circ$), in which case 
$U = 0$ and, hence, effectively $P = Q/I$. See Section~\ref{sec:results_inclination} for results with different inclinations.} 
\label{fig:phase_curves}
\end{figure}

Rayleigh scattering off gaseous atoms and molecules constitutes a dominant contribution to scattered light in 
cloud-free planetary atmospheres. Here, we compute the observables for Rayleigh scattering for both the scalar 
and vector cases, and we compare the results. Figure~\ref{fig:Ag} shows the geometric albedo ($A_g$) as a 
function of the single-scattering albedo for the two cases, and Figure~\ref{fig:phase_curves} shows the corresponding phase curves, $\Phi(\alpha)$. 

The geometric albedos for Rayleigh scattering using the scalar and vectorial treatments are not the same, 
as shown in Figure~\ref{fig:Ag}a. For conservative Rayleigh scattering ($\omega = 1.0$), i.e. a perfectly 
Rayleigh scattering atmosphere, the scalar phase function yields the commonly used value of $A_g = 0.75$. 
However, using the full phase matrix for the same case gives $A_g = 0.7977$, which is the more accurate value. 
The inset in Figure~\ref{fig:Ag} shows the fractional difference between $A_g$ in the two cases as a function of scattering albedo. Using the scalar phase function for computing $A_g$ in the present case yields values lower than the correct $A_g$ by up to 9\%, depending upon the scattering albedo. The differences between 
the two cases are lower for smaller $\omega$, i.e. for greater absorption. 

The phase curves for  both vector and scalar Rayleigh scattering for different values of $\omega$ are shown in the 
top panel of Figure~\ref{fig:phase_curves}. 
At large phase angles ($\alpha \gtrsim 110^{\circ}$), the phase curves for different $\omega$ are similar. However, for $\alpha \lesssim 110^{\circ}$, phase curves for lower $\omega$ have steeper gradients, for both the scalar and vector cases. For a given $\omega$, 
the differences between the phase curves for Rayleigh scattering obtained using the scalar 
and vector approaches are minimal. As shown in Figure~\ref{fig:phase_curves}, for low scattering albedos ($\omega \lesssim 0.5$) the phase curves obtained by the two approaches are nearly identical. As with $A_g$, the differences grow with $\omega$, but additionally the differences are now also a function of the phase angle. Even so, the 
maximum difference between the curves for $\omega = 1.0$ is only $\sim$5\%. Such differences in phase curves 
are not observable given the current precision of observations (Snellen et al. 2010; Demory et al. 2011). 

\begin{figure}[h]
\centering
\includegraphics[width = 0.5\textwidth]{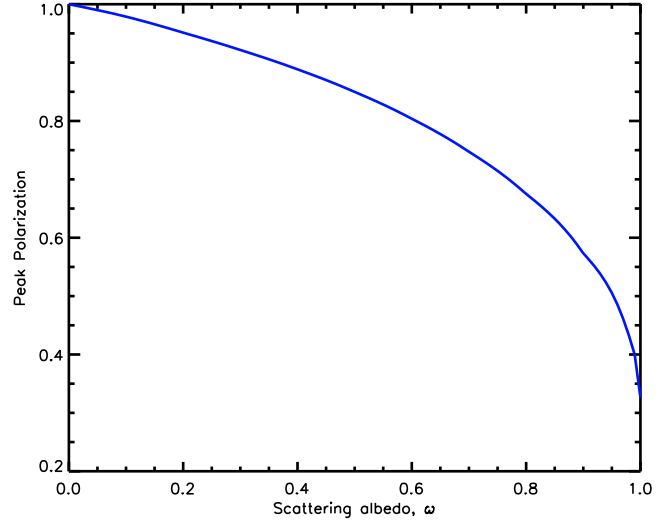}
\caption{Maximum polarization due to Rayleigh scattering as a function of the scattering albedo. The degree of polarization 
is given by $P = \sqrt{Q^2 + U^2}/I$ and is a function of the orbital phase. The vertical axis shows the maximum 
attainable P in an orbit for a given scattering albedo, shown as the horizontal axis. For circular edge-on orbits the peak polarization in attained close to quadrature phase, $\alpha \sim 90^\circ$.} 
\label{fig:peakpol}
\end{figure}

The degree of polarization of reflected light for a transiting planet with the Rayleigh phase matrix is shown in the lower panel of 
Fig.~\ref{fig:phase_curves}. The polarization curves, as a function of the phase angle, are 
shown for different $\omega$. For all $\omega$, the polarization peaks at $\alpha \sim 90^\circ$. 
However, as is well known in the literature (Buenzli \& Schmid 2009), the degree of polarization ($P$) 
increases with decreasing $\omega$. As shown in Fig.~\ref{fig:peakpol}, for highly absorptive 
atmospheres, say for $\omega = 0.1$, while the total intensity of reflected light is very low, the maximum 
polarization ($P_{max}$) is high, approaching 100\%. On the other hand, for highly scattering atmospheres, 
$P_{max}$ can be as low as 30\%, for $\omega = 1.0$. Since both the geometric albedo and the polarization 
curve for a given object at a given wavelength are related to the scattering albedo, a measurement of one can 
provide constraints on the other.

\begin{figure}[t]
\centering
\includegraphics[width = 0.5\textwidth]{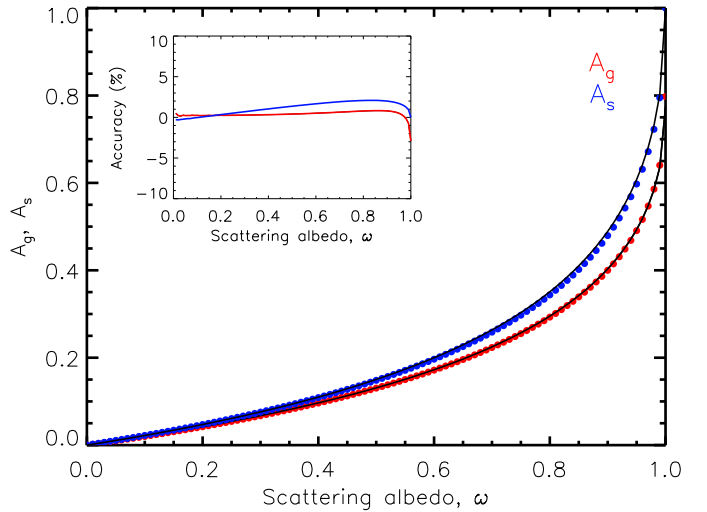}
\caption{Analytic fits for geometric and spherical albedos as a function of scattering albedo for 
conservative and non-conservative vector Rayleigh scattering. The red and blue circles show the 
geometric and spherical albedos, respectively. The black curves through the circles show the corresponding 
analytic fits, discussed in Section~\ref{sec:fits}. The inset shows the accuracy of the fits for each case.} 
\label{fig:fits}
\end{figure}

\subsection{Fits to Geometric and Spherical Albedos}
\label{sec:fits}

It is useful to quantify the dependence of the geometric and spherical albedos 
on the single-scattering albedo, so that given an observational albedo measurement 
one can constrain the underlying absorptive properties of the atmosphere. As is 
evident from Eq.~(\ref{eq:lambert}), for Lambert scattering, $A_g$ and $A_s$ are proportional to $\omega$. 
However, for other scattering phase functions, such as that of Rayleigh scattering, 
the dependence is non-linear. Even though $A_g$ and $A_s$ are expected 
to increase monotonically with $\omega$, the exact functional form can be 
non-trivial, but can be obtained by fits to numerical results. van de Hulst (1974) 
suggested the following functional form for $A_s$ for a semi-infinite atmosphere with a 
homogeneous cloud layer:
\begin{equation}
A_s = \frac{(1 - 0.139s)(1 - s)}{(1 + 1.170s)},  
\label{fig:hulst}
\end{equation}
where, $s = \sqrt{(1-\omega)/(1-\omega g)}$ and $g$ is the scattering asymmetry factor 
given by $<\cos\theta>$.

\begin{figure}[t]
\centering
\includegraphics[angle=-90, width = 0.5\textwidth]{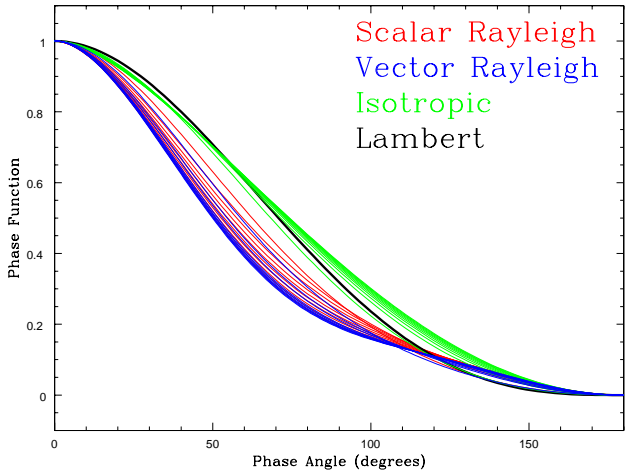}
\caption{Comparison of phase curves for different scattering phase functions. The phase curves 
for Rayleigh scattering (both scalar and vector) and isotropic scattering are shown for several $\omega$ values between 0 and 1; higher phase curves correspond to larger $\omega$. For Lambert scattering, the phase curve is independent of $\omega$.} 
\label{fig:pha_all}
\end{figure}

\begin{figure}
\centering
\includegraphics[width = 0.5\textwidth]{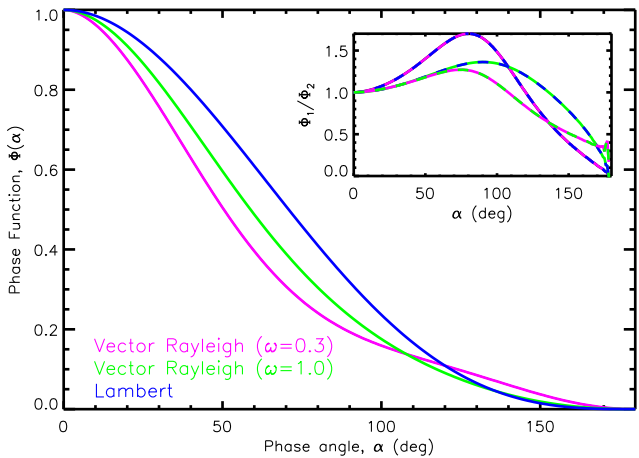}
\caption{Comparison between phase curves due to Rayleigh scattering (with vector formulation) and Lambert scattering. The solid curves in the main panel show three phase curves for the cases described in the legend. The multi-colored dashed curves in the inset show the ratios between pairs of solid curves, e.g. the magenta-green dashed curve is the ratio between the magenta and green solid curves.} 
\label{fig:phase_diff}
\end{figure}

\begin{figure*}
\centering
\includegraphics[width = \textwidth]{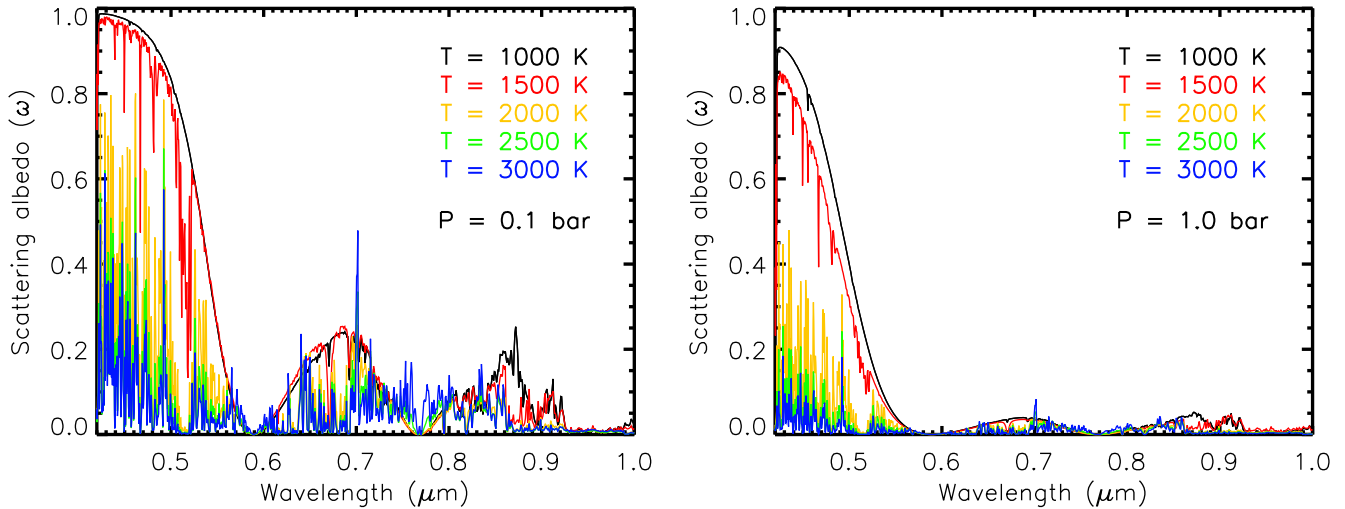}
\caption{Scattering albedos from model atmospheres. Scattering albedo spectra are shown for a representative range of pressures and temperatures observed in irradiated giant planet atmospheres. The models assume hydrogen-dominated atmospheres, with solar abundances, in chemical equilibrium. The scattering albedo is given by $\omega = \sigma_{scat}/(\sigma_{scat} + \sigma_{abs})$, where $\sigma_{scat}$ is the single-scattering cross section and $\sigma_{abs}$ is the absorption cross section. For wavelengths ($\lambda$) blueward of $\sim 0.55~\micron$, significant absorption due to Na, K, and molecular species cause low scattering albedos. For $\lambda \lesssim 0.55~\micron$, Rayleigh scattering becomes prominent, leading to high scattering albedos. See Section~\ref{sec:scat_albedos} for discussion.}
\label{fig:scat}
\end{figure*}

\begin{figure}
\centering
\includegraphics[width = 0.5\textwidth]{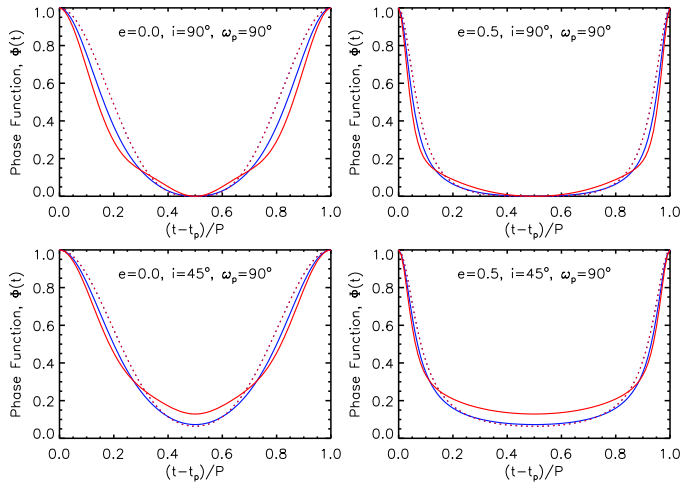}
\caption{Phase curves as a function of time from periastron passage for different orbital parameters. The $y$-axis shows the phase function. The orbital parameters are shown in the panels: eccentricity ($e$), inclination ($i$), and argument of periastron ($\omega_p$).  In each panel, the solid red and blue curves correspond to vector Rayleigh scattering with $\omega = 0.3$ and $\omega = 1.0$, respectively. The dotted red curve corresponds to Lambert scattering, which is independent of $\omega$.} 
\label{fig:phase_1}
\end{figure}

\begin{figure}
\centering
\includegraphics[width = 0.5\textwidth]{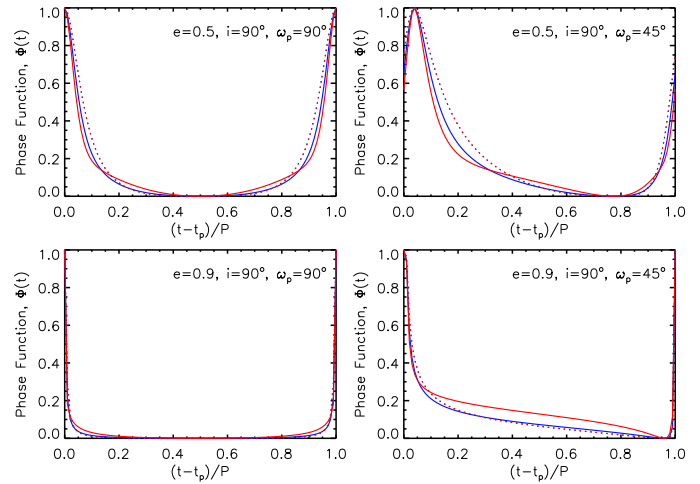}
\caption{Phase curves as a function of time from periastron passage. See Fig.~\ref{fig:phase_1} for description of axes. The orbital parameters for the curves are shown in each panel. The left and right panels in each row show the effect of the argument of periastron, for the same eccentricity and inclination. The top and bottom panels in each column show the effect of eccentricity, for the same argument of periastron and inclination. In each panel, the solid red and blue curves correspond to vector Rayleigh scattering with $\omega = 0.3$ and $\omega = 1.0$, respectively. The dotted red curve corresponds to Lambert scattering, which is independent of $\omega$.} 
\label{fig:phase_2}
\end{figure}

\begin{figure}
\centering
\includegraphics[width = 0.5\textwidth]{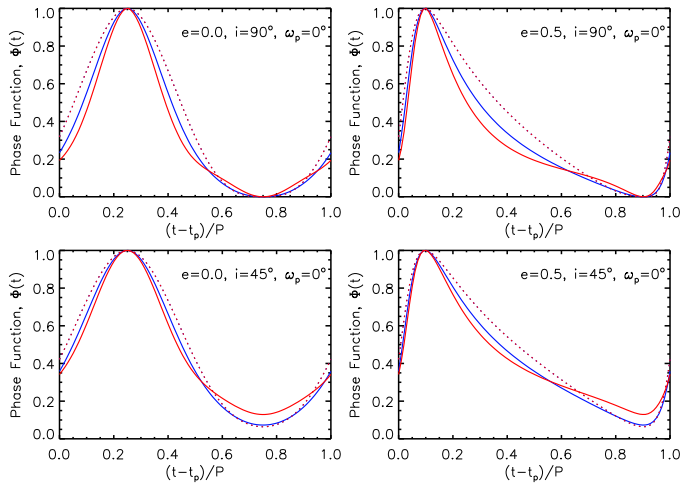}
\caption{Phase curves as a function of time from periastron passage for different orbital parameters. 
See Fig.~\ref{fig:phase_1} for description of the colors. The orbital parameters are shown in the panels.}
\label{fig:phase_3}
\end{figure}

\begin{figure*}
\centering
\includegraphics[width = \textwidth]{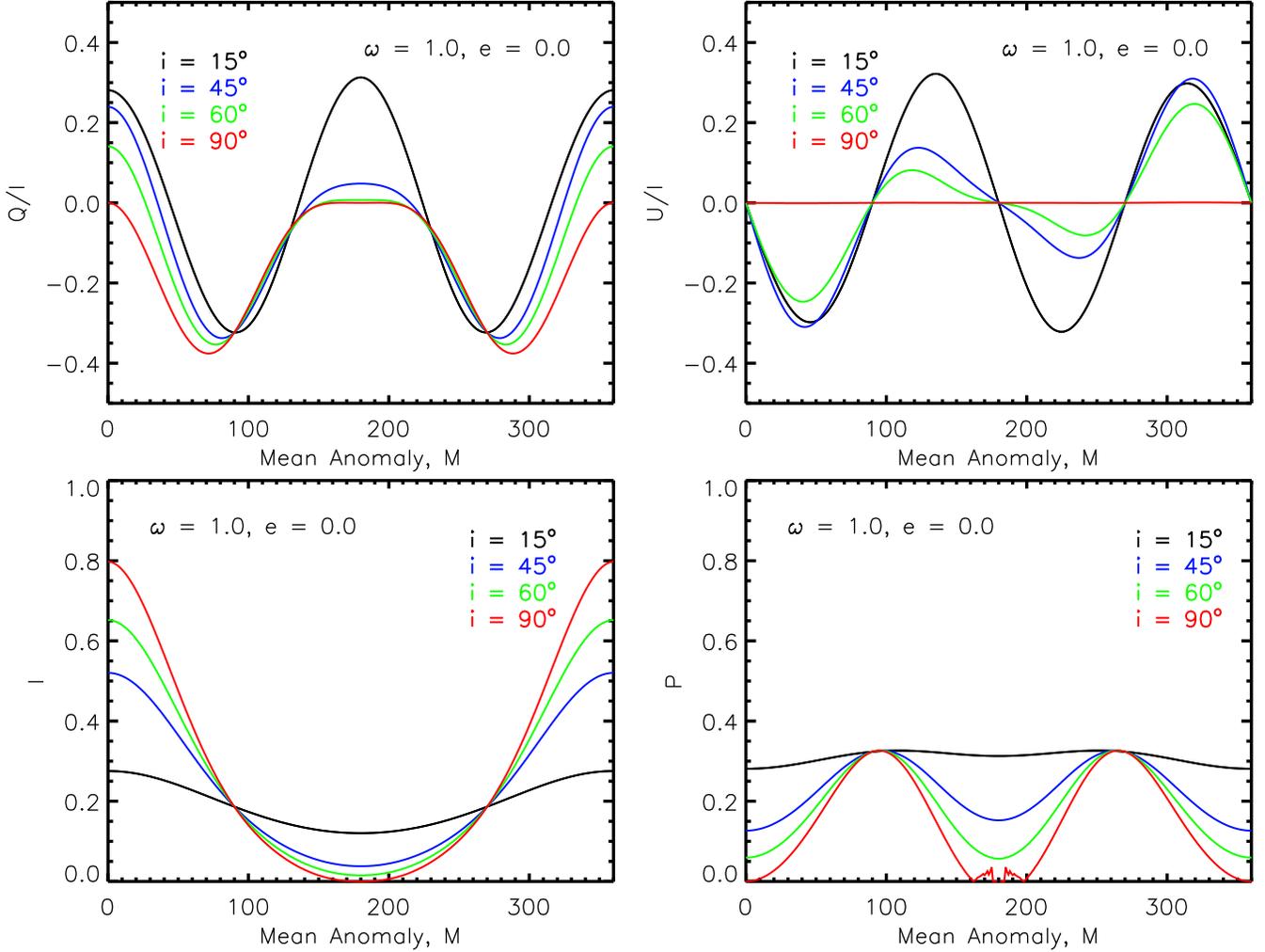}
\caption{Polarization curves for conservative Rayleigh scattering ($\omega = 1.0$) assuming circular orbits ($e = 0$). Various quantities involving the Stokes parameters (I, Q, U), integrated over the illuminated planetary surface, are shown as a function of the mean anomaly (M). The Stokes parameters are expressed as a fraction of the incident intensity. $I$ is the total intensity, and Q and U are the two  polarization parameters for linear polarization. The various quantities are shown for different orbital inclinations. For edge-on orbits ($i = 90^\circ$), the maximum value of $I$ in the orbit gives the geometry albedo, and U = 0 due to symmetry in the north-south direction. The degree of polarization, given by $P = \sqrt{Q^2 + U^2}/I$, and the polarization angle $\chi = 0.5 \tan^{-1}(U/Q)$ are also shown as a function of the mean anomaly. For the $i = 90^\circ$ case, the irregularities in the degree of polarization close to M = 180$^\circ$ are unphysical, caused numerically by division by vanishingly small values of I.} 
\label{fig:pol_circ}
\end{figure*}

\begin{figure*}
\centering
\includegraphics[width = \textwidth]{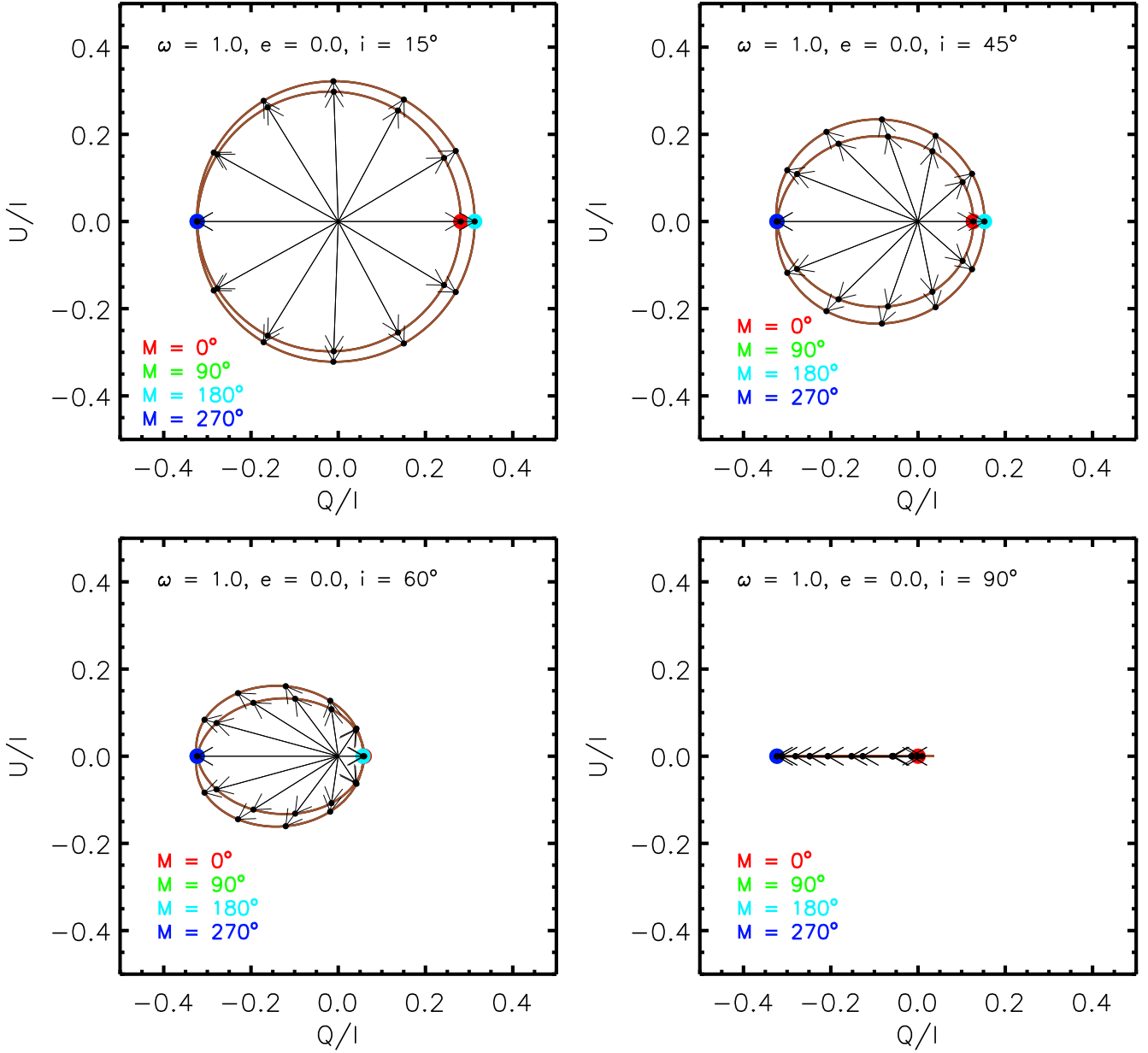}
\caption{Polarization curves for conservative Rayleigh scattering ($\omega = 1.0$) assuming circular orbits ($e = 0$). The trajectory of the planet in the Q/I -- U/I plane is shown for different inclinations. The arrows show the 
position of the planet in the Q/I -- U/I plane at different mean anomalies (M), spaced apart by 15$^o$. Sample 
locations for M of 0$^o$, 90$^o$, 180$^o$, and 270$^o$, are shown by the colored circles described in the 
legend. See section~\ref{sec:results_inclination} for discussion.} 
\label{fig:pol_circ_1}
\end{figure*}

\begin{figure*}
\centering
\includegraphics[width = \textwidth]{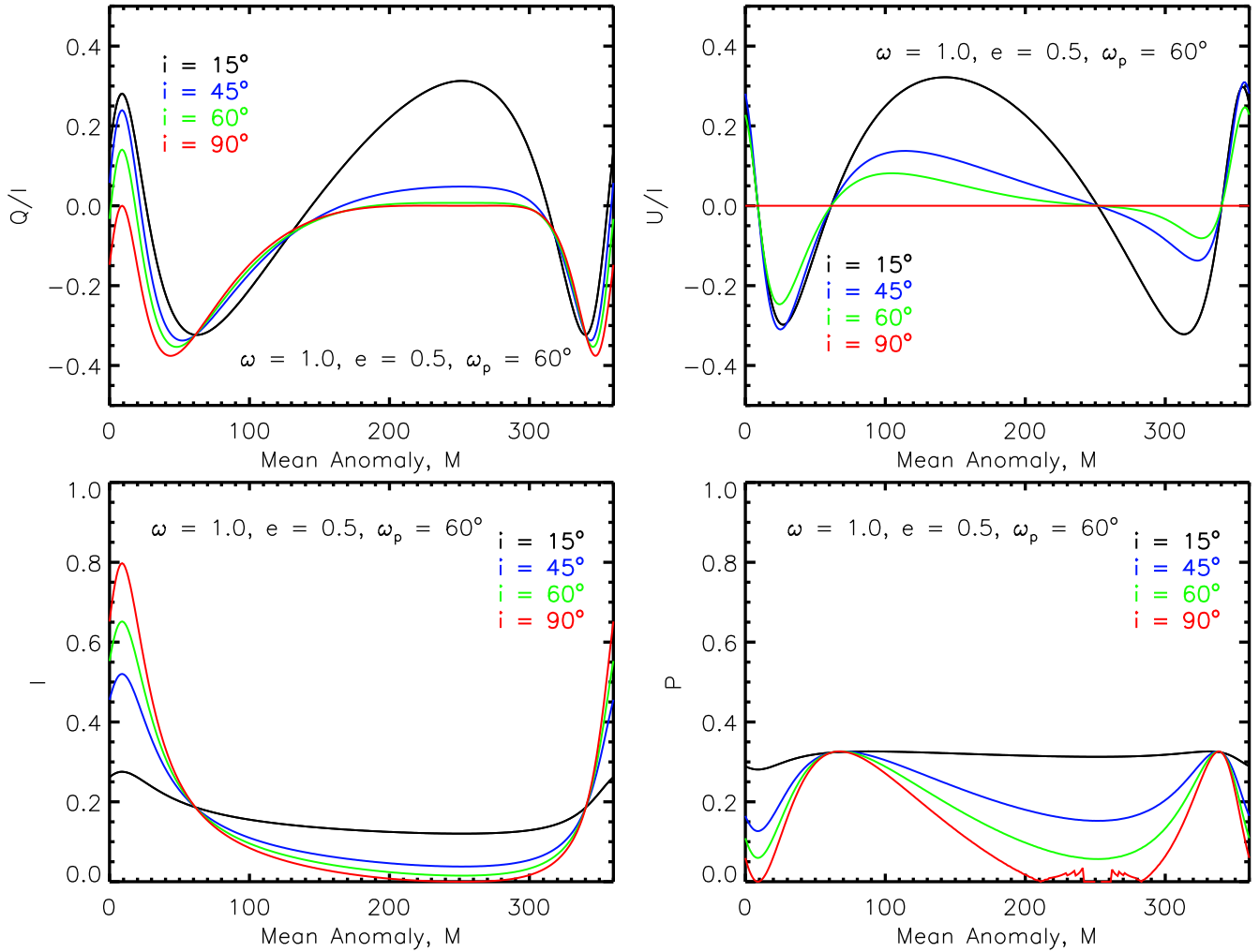}
\caption{Polarization curves for conservative Rayleigh scattering for eccentric orbits. See description in 
Fig.~\ref{fig:pol_circ}. For illustration, an eccentricity ($e$) of 0.5 and an argument of periastron ($\omega_p$) of 60$^\circ$ are assumed. The longitude of ascending node ($\Omega$) is assumed to be 90$^\circ$. For this eccentric case, the Stokes parameters vary asymmetrically with the mean anomaly, contrary to the circular case. However, for edge-on orbits the Stokes U still vanishes.} 
\label{fig:pol_ecc}
\end{figure*}

\begin{figure*}
\centering
\includegraphics[width = \textwidth]{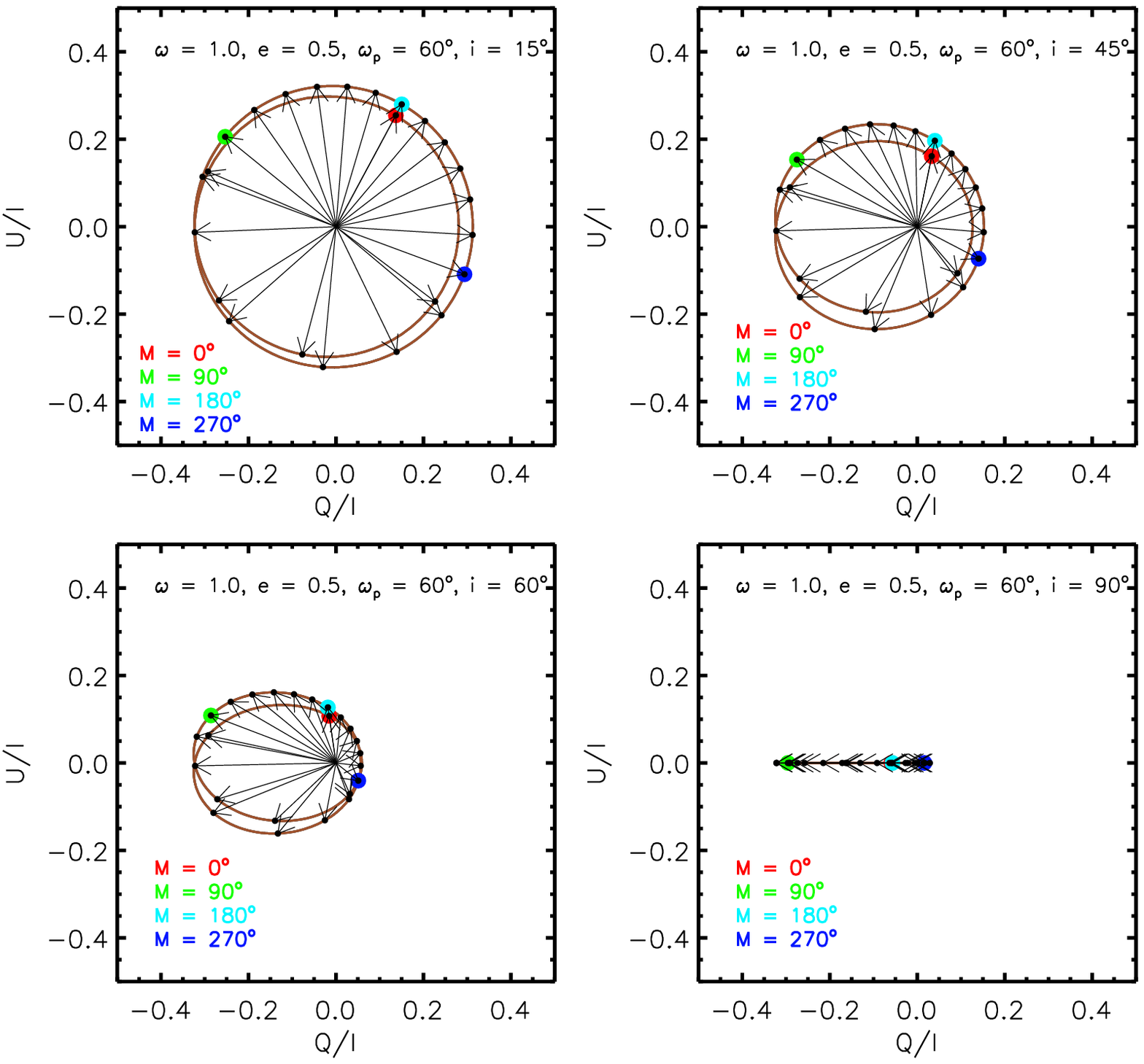}
\caption{Polarization curves for conservative Rayleigh scattering for eccentric orbits. The trajectories in the Q/I -- U/I plane for eccentric orbits are rotated with respect to those for circular orbits for the corresponding inclinations 
shown in Fig.~\ref{fig:pol_circ_1}. The arrows mark the mean anomalies (M) in the orbit, spaced apart by 15$^o$. The locations for M of 0$^o$, 90$^o$, 180$^o$, and 270$^o$, are shown by the colored circles described in the 
legend. The orbital parameters for each case are shown in the panels.} 
\label{fig:pol_ecc_1}
\end{figure*}

Inspired by Eq.(\ref{fig:hulst}), we obtain fits to $A_g$ and $A_s$ for non-conservative Rayleigh 
scattering in semi-infinite atmospheres, using the vector formulation, as follows:
\begin{equation}
A_g = 0.7977\frac{(1 - 0.23s)(1 - s)}{(1 + 0.72s)(0.95 + 0.08\omega)} \label{eq:ag_fit}
\end{equation}
and
\begin{equation}
A_s = \frac{(1 - 0.15s)(1 - s)}{(1 + 1.05s)},
\end{equation}
where $s = \sqrt{1-\omega}$. As shown in Fig.~\ref{fig:fits}, the fit for $A_g$ is accurate to 
within 1\% for $\omega \lesssim 0.99$, and to within $\sim$ 3\% for $\omega \sim 1$. 
For $\omega = 1.0$, a better fit is obtained for $A_g$ without the term $(0.95 + 0.08\omega)$ 
in the denominator in Eq.(\ref{eq:ag_fit}). The fit for $A_s$ is accurate to within 2\% for all 
$\omega$. 

The above expressions can be used to retrieve the scattering albedo from an 
observed geometric albedo, under the assumption of a homogenous, semi-infinite 
Rayleigh scattering atmosphere. The $\omega$ thus obtained is a representative 
average scattering albedo at a given wavelength of the observation. 

\subsection{Comparison between different phase curves}

It is instructive to compare the phase curves resulting from different single-scattering 
phase functions. Figure~\ref{fig:pha_all} shows the phase curves for 
scattering due to Rayleigh scattering, isotropic scattering, and Lambert scattering. 
While the Lambert phase curve is independent of $\omega$, the isotropic and 
Rayleigh phase curves vary with $\omega$. We find that the Lambertian and isotropic 
phase curves predict up to a factor of $\sim$2 higher phase functions compared to the 
Rayleigh phase curves for all $\omega$ for phase angles below $\sim$120$^{\circ}$, 
with the maximum differences occurring for $\alpha \sim 70^\circ - 90^\circ$. Furthermore, 
for Rayleigh scattering, higher $\omega$ leads to higher phase functions for most 
phase angles. As shown in Figure~\ref{fig:phase_diff}, an $\omega = 1.0$ results in up to 
a factor of $\sim$1.5 higher phase function compared to a phase curve with $\omega = 0.3$, 
and up to $\sim$1.3 higher phase function compared to a Lambert phase curve.
Consequently, observations at phase angles corresponding to maximum differences 
in the phase curves can potentially be used to constrain scattering phenomena in planetary 
atmospheres. The magnitude of differences between the phase curves due 
to the different scattering phenomena are comparable to the those between phase curves 
of planets with different atmospheric and orbital characteristics. For example, Sudarsky et al. (2005) 
reported theoretical phase curves of extrasolar giant planets as a function of orbital distance, 
between which typical differences are comparable to those seen in the phase curves in 
Fig.~\ref{fig:pha_all} and Fig.~\ref{fig:phase_diff}.    

\subsection{Scattering Albedos from Model Atmospheres}
\label{sec:scat_albedos}
As discussed in previous sections, the geometric albedos, phase curves, and polarization of scattered light 
depend strongly on the scattering albedo. The scattering albedo is given by, $\omega = \sigma_{scat}/(\sigma_{scat} + \sigma_{abs})$, where $\sigma_{scat}$ is the single-scattering cross section and  $\sigma_{abs}$ is the 
absorption cross section. As such, the scattering albedo is a strong function of wavelength ($\lambda$), 
and of the atmospheric parameters such as temperature ($T$), pressure ($P$), and composition, all of which 
govern $\sigma_{scat}$ and $\sigma_{abs}$. Figure~\ref{fig:scat} shows scattering albedos in the visible wavelengths, for a wide range of $T$, and for $P$ of 0.1 and 1 bar which are representative of the pressure layers probed by visible/near-IR observations of exoplanetary atmospheres (Burrows et al.~2008; Sharp \& Burrows 2007; Madhusudhan et al.~2011a,b). Here, we assume cloud-free, solar-abundance atmospheres in chemical equilibrium, at the given $P$ and $T$. 

As shown in Fig.~\ref{fig:scat}, the scattering albedos are significant at short wavelengths ($\lambda \lesssim 0.55$), and decline steeply to very low values ($\omega \lesssim 0.2$) for $\lambda \gtrsim 0.55$. The increase in $\omega$ with decreasing $\lambda$ is due primarily to Rayleigh scattering. Furthermore, $\omega$ is a strong function of temperature. Extremely irradiated atmospheres with $T \gtrsim 2000$ K provide substantial molecular absorption, leading to low scattering albedos across all $\lambda$. The lower temperature end of irradiated atmospheres, with representative temperatures of 1000 - 1500 K has the highest $\omega$ at short wavelengths. Finally, $\omega$ also depends on the pressure, and is higher for $P = 0.1$ bar than for $P = 1$ bar, and is generally higher for lower pressures. Consequently, an average scattering albedo derived from data in a given spectral  bandpass is indicative of the chemical and thermal properties of the atmosphere probed by the observations.

\subsection{Time Dependence}
\label{sec:time_dep}
In previous sections, we expressed the phase curves for different scattering phenomena as functions of the phase angle ($\alpha$). However, an observed phase curve, as a function of time, also depends on the orbital parameters, as discussed in Section~\ref{sec:orb_phase}. Phase curves due to Rayleigh scattering and Lambert scattering as a function of time from periastron passage are shown in Figs.~\ref{fig:phase_1}-\ref{fig:phase_3}, 
for different orbital parameters. For Rayleigh scattering, we consider two examples corresponding to 
$\omega = 1.0$ (conservative scattering) and $\omega = 0.3$ (non-conservative scattering). The Lambert 
phase curve is independent of $\omega$.  In all cases, the peak of the phase curve corresponds to a 
phase angle of 0$^o$, i.e. full illumination at opposition, and the minimum corresponds to a phase angle of $180^o$. Here, we assume the longitude of the ascending node ($\Omega$) to be $90^\circ$.

The differences between the phase curves due to the different scattering phase functions can be used 
to distinguish between the underlying scattering phenomena, depending upon the orbital parameters. 
The phase curves for vector Rayleigh scattering generally have steeper gradients than those for Lambert scattering. 
However, the absolute differences between the curves vary with the orbital phase, and are a strong function 
of the orbital parameters. For example, for the case of $e=0, i = 90^o, \omega_p = 90^o$ in Fig.~\ref{fig:phase_1} (top-left panel), the planet-star flux contrast for Lambert scattering with $\omega = 0.3$ can be up to a factor of 2 higher than that 
for the corresponding Rayleigh scattering case. In this case, the maximum difference occurs at an orbital phase 
corresponding to (t-t$_{\rm p}$)/P $\sim$ 0.2. It follows that observations made around this phase, in 
comparison with those made near the peak of the phase curve, could provide a constraint on the scattering mechanism. Thus, given the orbital parameters of a planet, similar times can be predicted  for orbital phases which could provide the best diagnostics between the different scattering phenomena. However, in some cases, such as in the top-right panel of Fig.~\ref{fig:phase_2}, the differences between the different phase curves are only marginal. 

\section{Polarization as a Function of Orbital Parameters}
\label{sec:results_inclination}

The polarization curves in Section~\ref{sec:results_rayleigh} assumed the planetary orbit to be edge-on, namely that the inclination angle ($i$) was 90$^\circ$. Such an assumption is reasonable for planets in the solar system and for transiting extrasolar planets. In this section, we relax that assumption and study the dependence of the polarization on the orbital parameters: inclination ($i$), eccentricity ($e$) and the argument of periastron ($\omega_p$). For the case of edge-on orbits, the Stokes parameter U is zero due to the latitudinal symmetry of 
the illuminated surface of the planet, whereas the Stokes parameter Q is non-zero due to asymmetry in the 
longitudinal direction. For an orbit that is not edge-on, however, both Q and U are non-zero, leading to a non-zero polarization angle, $\chi = \frac{1}{2} \tan^{-1}(U/Q)$.  

The polarization parameters provide strong constraints on the inclination of the planetary orbit, as is well 
known in the literature on binary stars (e.g. Rudy and Kemp 1978; Schmid 1992; Harries and Howarth 1996). 
Recently, measurements of polarization parameters are also being attempted for giant exoplanetary 
atmospheres (Berdyugina 2008, 2011a,b; but cf Wiktorowicz 2009). Here, we explore the dependence of the 
polarization parameters on the orbital properties of a planetary system using our model formalism developed in previous sections. 

The time-dependent Stokes parameters in the observer's frame can be derived using the following 
prescription (also see Schmid 1992): 
\begin{enumerate}
\item{Given a mean anomaly ($M$), the true anomaly ($\theta$) can be obtained by solving the 
Kepler's equation, Eq. (\ref{eq:kepler}).}
\item{The effective phase angle corresponding to a true anomaly of $\theta$ in an inclined orbit is 
given by 
\begin{equation}
\alpha = \sin^{-1}\sqrt{\sin^2\eta \cos^2 i +  \cos^2\eta}~; 0 \leq \eta \leq \pi
\end{equation}
and 
\begin{equation}
\alpha = \pi - \sin^{-1}\sqrt{\sin^2\eta \cos^2 i +  \cos^2\eta}~; \pi < \eta < 2\pi
\label{eq:incl_1}
\end{equation}
where, $\eta$ = ($\theta$ + $\omega_p$) is the true longitude, $\omega_p$ is the argument of periastron, 
and $i$ is the orbital inclination.}  
\item{Given the $\alpha$, the Stokes vector ({\textbf S}) in the plane of the orbit is computed using the 
prescription in Section~\ref{sec:results}.} 
\item{Finally, the Stokes vector ({\textbf S}) must be rotated into the frame of reference of the observer. 
The rotation angle is given by: 
\begin{equation}
\gamma = \tan^{-1}[\cot(\eta)/\cos(i)].
\label{eq:incl_2}
\end{equation}}
\end{enumerate}
Here, we have assumed the longitude of the ascending node ($\Omega$) to be $\pi/2$. The observed Stokes 
vector (S$^\prime$) is obtained by ${\textbf S^\prime = \textbf{L}\cdot\textbf{S}}$, 
where ${\textbf L}$ is the transformation matrix given by Eq.(\ref{eq:stokes_transform}). 

Model polarization curves as a function of the orbital inclination for circular orbits are shown in Figs.~\ref{fig:pol_circ} \& \ref{fig:pol_circ_1}. Here, we assume conservative Rayleigh scattering with the vector phase matrix discussed in Section~\ref{sec:results_rayleigh}. The Stokes parameters (I, Q, and U), the degree of polarization ($P = \sqrt{(Q^2 + U^2)}/I$), and the polarization angle ($\chi = \frac{1}{2} \tan^{-1}(U/Q)$) are shown as a function of the mean anomaly (M) for different inclinations. For edge-on orbits ($i = 90^\circ$), the Stokes U parameter vanishes at all orbital phases, due to symmetry in the illuminated planetary area in the north-south direction, whereas the Stokes Q is non-zero. For all other inclinations ($i < 90^\circ$), both Q and U are non-zero. On the other hand, the planetary phase curve (i.e. Stokes I as a function of M) is steepest for edge-on orbits and gradually flattens out for smaller inclinations, leading to a uniform phase curve for a face-on orbit ($i = 0^\circ$). Secondly, while the maxima of Q/I and U/I in the orbit increase with decreasing inclination, the minima remain relatively unchanged. 

The variation of the degree of polarization ($P = \sqrt{(Q^2 + U^2)}/I$) as a function of M for different orbital 
inclinations is shown in the lower-right panel of Fig.~\ref{fig:pol_circ}. A face-on orbit yields a constant polarization 
at all orbital phases, and an edge-on orbit causes the maximum peak-to-trough variation with orbital phase. Strategic 
monitoring of $P$ over the orbit provides useful diagnostics on the orbital inclination of the system. The peak polarization ($P_{\rm max}$) occurs at the two quadrature phases, mean anomalies ($M$) 
of  $90^\circ$ and $270^\circ$, independent of the inclination. In addition, $P_{\rm max}$ is identical for all 
inclinations. On the other hand, the minimum polarization ($P_{\rm min}$) is a strong function of the inclination, 
and ranges between $P_{\rm max}$ for $i = 0^\circ$ and 0 for $i = 90^\circ$. For circular orbits, $P_{\rm min}$ 
occurs at the conjunctions, i.e. orbital phases of 0 and $180^\circ$. Consequently, the inclination of a planetary 
orbit can be constrained by monitoring the extrema in the polarization phase curve of the planet. 

The influence of the orbital inclination on the polarization parameters is also apparent in the trajectory of the planet in Q -- U space. Fig.~\ref{fig:pol_circ_1} shows the planetary orbits in the Q/I -- U/I space for different inclinations. For each 
inclination, the trajectory is a double loop structure which becomes increasingly eccentric with increasing inclination. For face-on orbits a double-circular trajectory is obtained, whereas for perfectly edge-on orbits a horizontal trajectory  (with U = 0) is obtained. Consequently, the ellipticity of a planet's path in the Q/I -- U/I space can be used to infer the orbital inclination of the system. The arrows in each panel show the variation of the polarization angle with the mean anomaly, M. The colored circles show the position of the planet in Q/I -- U/I space, and hence its polarization angle ($\chi = \frac{1}{2} \tan^{-1}(U/Q)$), at four different times (denoted by the mean anomaly, M) in the orbit. The arrows 
show the same at $15^\circ$ intervals between M of $0^\circ$ and $360^\circ$. 

The polarization parameters for eccentric orbits for different orbital inclinations are shown in Figs.~\ref{fig:pol_ecc} \& \ref{fig:pol_ecc_1}. For purposes of illustration, we consider an orbital eccentricity ($e$) of 0.5, and an argument of periastron ($\omega_p$) of 60$^\circ$. Again, we set the longitude of the ascending node ($\Omega$) to 90$^\circ$. The variation of the total intensity ($I$) with the mean anomaly ($M$) is governed by the Kepler equation, Eq.~\ref{eq:kepler}, which relates $M$ to the phase angle ($\alpha$), depending on $e$ and $\omega$. Contrary to the circular case, the variation in the polarization parameters is asymmetric in time for eccentric orbits. However, similar to the case  of circular orbits, the mean anomalies corresponding to the maxima and minima remain largely unchanged with the 
orbital inclination. Also similar to the circular case, the peak polarization in eccentric orbits is independent of the orbital inclination ($i$), whereas the minimum polarization is a strong function of $i$. 

The trajectory of an eccentric orbit in the Q/I -- U/I space is shown in Fig.~\ref{fig:pol_ecc_1}. As in the circular case, the orbit traces a double loop structure which becomes increasingly eccentric with increasing inclination. However, for the eccentric case the polarization angles are naturally shifted in time, as shown by the positions of the planet in the Q/I -- U/I plane at the different mean anomalies. Consequently, by observing a polarimetric orbit of a planet in the Q -- U space, the inclination as well as the eccentricity and the argument of periastron of the planetary orbit can, in principle, be determined. 

\section{Discussion and Summary}
\label{sec:summary}

Observations are becoming increasingly capable of detecting geometric albedos, 
phase curves, and polarization from extrasolar planets in reflected light. We have provided an 
integrated analytic framework to interpret such observations. Our approach makes accessible a 
computationally efficient procedure with which to estimate an average scattering albedo ($\omega$) 
from an observable in reflected light, such as a geometric albedo or a phase curve. 
We follow the H-function approach of Chandrasekhar (1950,1960) which has 
been used in subsequent studies to derive analytic solutions for scattered emergent flux for various 
scattering phenomena. We consolidate the formalism, solution techniques, and results from analytic models 
available in the literature, but often scattered in various papers, and present a systematic procedure to compute albedos, phase curves, and polarization of reflected light. 

In this work, we considered cloud-free, homogeneous, semi-infinite atmospheres scattering in accordance with different scattering phase functions. We consider both conservative ($\omega = 1.0$) and non-conservative ($\omega < 1.0$) scattering. We compute geometric albedos and phase curves for Rayleigh scattering, with scalar and vector phase functions, isotropic scattering, asymmetric scattering, and Lambert scattering. We also compute polarization curves for the case of vector Rayleigh scattering, and provide a step-by-step procedure to obtain the disk-integrated emergent flux and polarization, for a given scattering albedo and phase angle, with the appropriate angular transformations to the celestial reference frame. The phase curves and polarization curves are also computed as a function of the mean anomaly, or a time coordinate, for given orbital parameters. 

Rayleigh scattering is a dominant scattering mechanism in cloud-free planetary atmospheres. 
An accurate calculation of the geometric albedo ($A_g$) for Rayleigh scattering must 
include the effect of polarization, using the full Rayleigh phase matrix for single-scattering, 
contrary to the customarily adopted scalar phase function. For, conservative Rayleigh scattering 
($\omega = 1.0$), $A_g = 0.7977$, as opposed to the familiar value of 0.75 obtained using the scalar treatment. 
The scalar formulation leads to geometric albedos that are lower than the actual values by up to 9\%, 
depending on $\omega$. We compute albedos and phase curves for Rayleigh scattering for a range of 
$\omega$, and report analytic fits to the geometric and spherical albedos as a function of $\omega$. 
Based on the polarization curves, we also compute the peak polarization due to Rayleigh scattering 
as a function of $\omega$. The emergent intensity and degree of polarization of reflected light for 
Rayleigh scattering depend strongly on $\omega$. The emergent intensity at full phase, and, hence, the 
geometric albedo, both increase with $\omega$. On the other hand, the degree of polarization ($P$) decreases 
with increasing $\omega$. 

Motivated by recent efforts to measure polarization of reflected light from exoplanets, we compute 
polarization curves for homogeneous Rayleigh scattering atmospheres over a wide range of orbital 
parameters using the vector Rayleigh phase matrix. We explore the dependence of the Stokes parameters 
on the orbital inclination and eccentricity. We demonstrate how the inclination of a planetary orbit can be 
constrained by monitoring the extrema in the polarization phase curve of the planet, as well as the ellipticity 
of the orbit in the Q-U plane.

We summarize the results of our work as follows: 
\begin{itemize}
\item{We present an analytic framework to interpret observables of reflected light (phase curves, geometric albedos, and polarization parameters) from extrasolar planets. Our analytic models assume cloud-free homogeneous and semi-infinite atmospheres, and we consider several different scattering phase functions.} 
\item{We show that observations of phase curves of exoplanets in reflected light can be used to constrain the underlying scattering mechanisms (e.g. Lambert versus Rayleigh, etc.) in their atmospheres.} 
\item{We show that observed geometric albedos of exoplanetary atmospheres can be used to constrain their average 
single scattering albedos, which are indicative of their chemical and thermal properties, assuming different scattering 
mechanisms.}
\item{We demonstrate, using the vector Rayleigh phase matrix, how polarization curves can be used to constrain orbital parameters (inclinations and eccentricities) of exoplanets.}
\end{itemize}

The simplification of an analytic approach does introduce some natural caveats. While our requirement of semi-infinite atmospheres is reasonable for all gaseous atmospheres, from super-Earths to giant planets, our assumption of homogeneity is meant to represent only an average scattering albedo in the atmosphere in the observed spectral/photometric bandpass. In general, the scattering albedo in a gaseous atmosphere would not remain constant at all altitudes and wavelengths. Our approach is aimed at providing a simple and efficient procedure to interpret reflected light observations of exoplanets which are currently limited by sparse data.

Detailed numerical models exist in the literature which solve the full radiative transfer problem with an inhomogeneous atmosphere in plane-parallel atmospheres. However, given the large number of free parameters and computational complexity of such models, it is generally not feasible to use them for formally fitting and retrieving atmospheric parameters from the limited data in reflected light. The analytic tools provided in our work makes it feasible to derive averaged atmospheric properties, such as an average scattering albedo or a representative scattering phase function, from the data. Our code can be used in conjunction with standard parameter  estimation methods to retrieve average scattering albedos from observed geometric albedos, as well as orbital parameters from observed phase curves and polarization curves of extrasolar planets. Our code is freely available on request by email to either of the authors.

\acknowledgements{The authors acknowledge support in part under NASA ATP grant
NNX07AG80G, HST grants HST-GO-12181.04-A and HST-GO-12314.03-A, and
JPL/Spitzer Agreements 1417122, 1348668, 1371432, and 1377197. We thank 
J. Bjorkman and K. Bjorkman for helpful discussions.}  
\newline

\end{document}